\documentclass[conference]{IEEEtran}
\IEEEoverridecommandlockouts
\usepackage{acronym}
\acrodef{pdd}[PDD]{penalty dual decomposition}
\acrodef{admm}[ADMM]{alternating direction method of multipliers}
\acrodef{ps}[PS]{phase shifter}
\acrodef{rfc}[RFC]{radio-frequency chain}
\acrodef{isl}[ISL]{inter-satellite link}
\acrodef{leo}[LEO]{low Earth orbit}
\acrodef{meo}[MEO]{medium Earth orbit}
\acrodef{geo}[GEO]{geostationary Earth orbit}
\acrodef{ut}[UT]{user terminal}
\acrodef{upa}[UPA]{uniform planar array}
\acrodef{csi}[CSI]{channel state information}
\acrodef{ofdm}[OFDM]{orthogonal frequency division multiplexing}
\acrodef{los}[LOS]{line-of-sight}
\acrodef{toa}[TOA]{time-of-arrival}
\acrodef{pace}[PACE]{positioning-aided channel estimation}
\acrodef{mcrb}[MCRB]{misspecified Cramér-Rao bound}
\acrodef{awgn}[AWGN]{additive white Gaussian noise}
\acrodef{crb}[CRB]{Cramér-Rao bound}
\acrodef{cfo}[CFO]{carrier frequency offset}
\acrodef{peb}[PEB]{position error bound}
\acrodef{crb}[CRB]{Cramér-Rao bound}
\acrodef{lb}[LB]{lower bound}
\acrodef{rmse}[RMSE]{root mean squared error}
\acrodef{fim}[FIM]{Fisher information matrix}
\acrodef{tdd}[TDD]{time division duplex} 
\acrodef{wmmse}[WMMSE]{weighted minimal mean squared error} 
\acrodef{qcqp}[QCQP]{quadratically constrained quadratic program}
\acrodef{mrt}[MRT]{maximum ratio transmission}
\acrodef{zf}[ZF]{zero-forcing}
\acrodef{gnss}[GNSS]{global navigation satellite system}
\acrodef{itu}[ITU]{International Telecommunication Union}
\acrodef{pab}[PAB]{Position-Aided Beamforming}
\acrodef{vdb}[VDB]{Vertically Directed Beamforming}
\acrodef{bs}[BS]{base station}
\acrodef{ntn}[NTN]{non-terrestrial networks}
\acrodef{uav}[UAV]{unmanned aerial vehicle}
\acrodef{tdma}[TDMA]{time-division multiple access}
\acrodef{haps}[HAPS]{high-altitude platform stations}
\acrodef{6g}[6G]{the sixth generation}
\acrodef{bse}[BSE]{beam squint effect}
\acrodef{cp}[CP]{cyclic prefix}
\acrodef{cpu}[CPU]{central processing unit}
\acrodef{dof}[DOF]{degrees-of-freedom}
\acrodef{elaa}[ELAA]{extremely large antenna array}
\acrodef{ff}[FF]{far-field}
\acrodef{las}[L\&S]{localization and sensing}
\acrodef{nf}[NF]{near-field}
\acrodef{ris}[RIS]{reconfigurable intelligent surface}
\acrodef{rtt}[RTT]{round-trip-time}
\acrodef{sinr}[SINR]{signal-to-interference-plus-noise ratio}
\acrodef{sns}[SNS]{spatial non-stationarity}
\acrodef{swm}[SWM]{spherical wave model}

\acrodef{siso}[SISO]{single-input-single-output}
\acrodef{mimo}[MIMO]{multi-input-multi-output}
\acrodef{ue}[UE]{user equipment}
\acrodef{dmimo}[D-MIMO]{distributed MIMO}
\acrodef{sp}[SP]{scatter point}
\acrodef{nlos}[NLOS]{non-line-of-sight}
\acrodef{tdoa}[TDOA]{time-difference-of-arrival}
\acrodef{am}[AM]{artificial multipath}
\acrodef{an}[AN]{artificial noise}
\acrodef{psd}[PSD]{power spectral density}
\acrodef{pdf}[PDF]{probability distribution function}
\acrodef{aoa}[AOA]{angle-of-arrival}
\acrodef{aod}[AOD]{angle-of-departure}
\acrodef{moo}[MOO]{multi-objective optimization}
\acrodef{qos}[QoS]{quality of service}
\acrodef{sdp}[SDP]{semi-definite programming}
\acrodef{lmi}[LMI]{linear matrix inequality}
\acrodef{sdr}[SDR]{semi-definite relaxation}
\acrodef{rcs}[RCS]{radar cross section}
\acrodef{isac}[ISAC]{integrated sensing and communication}
\acrodef{pdd}[PDD]{penalty dual decomposition}
\acrodef{bcd}[BCD]{block coordinate descent}
\usepackage{color}

\usepackage{pgfplots}
\usepackage{tikz}
\usetikzlibrary{calc}
\makeatletter
\newcommand{\gettikzxy}[3]{%
  \tikz@scan@one@point\pgfutil@firstofone#1\relax
  \edef#2{\the\pgf@x}%
  \edef#3{\the\pgf@y}%
}
\usetikzlibrary{spy,backgrounds}
\usepackage{mathrsfs}
\usepackage{booktabs} 
\usepackage{amsmath}
\usepackage{graphicx} 
\usepackage{epstopdf}
\usepackage{amssymb}
\usepackage{amsfonts}
\usepackage{amsthm}
\usepackage{cite}
\usepackage{bm,comment}
\usepackage{algorithm}
\usepackage{algpseudocode}

\usepackage{subeqnarray}
\usepackage{subfigure}
\usepackage{multicol}
\usepackage{multirow}
\usepackage{diagbox}
\usepackage{slashbox}
\usepackage{stfloats}
\usepackage{float}
\usepackage{color} 
\usepackage{cases}
\usepackage{lipsum}

\newtheorem{remark}{\bf{Remark}}

\usepackage[bookmarks,colorlinks]{hyperref} 
\hypersetup{colorlinks,citecolor= red,filecolor= blue,linkcolor= blue,urlcolor=blue}
\allowdisplaybreaks
\begin{document}
\setlength{\textfloatsep}{4pt}

\bstctlcite{IEEEexample:BSTcontrol}
\title{Enabling Scalable Distributed Beamforming via Networked LEO Satellites Towards 6G}
\author{
Yuchen Zhang, \emph{Member, IEEE} and Tareq Y. Al-Naffouri, \emph{Fellow, IEEE}
\thanks{
This publication is based upon work supported by King Abdullah University of Science and Technology (KAUST) under Award No. ORFS-CRG12-2024-6478 and Global Fellowship Program under Award No. RFS-2025-6844.

Yuchen Zhang and Tareq Y. Al-Naffouri are with the Electrical and Computer Engineering Program, Computer, Electrical and Mathematical Sciences and Engineering (CEMSE), King Abdullah University of Science and Technology (KAUST), Thuwal 23955-6900, Kingdom of Saudi Arabia (e-mail: \{yuchen.zhang; tareq.alnaffouri\}@kaust.edu.sa).
}}
\maketitle

\begin{abstract}
In this paper, we propose scalable distributed beamforming schemes over networked low Earth orbit (LEO) satellite systems that rely solely on statistical channel state information (CSI). We begin by introducing the LEO satellite network system model and presenting pragmatic yet effective analog beamformer and user-scheduling designs. 
We then derive a closed-form lower bound on the ergodic sum rate, based on the hardening bound, using which we formulate a per-satellite power-constrained sum rate maximization problem for the digital beamformer design. Next, we provide a centralized solution, obtained via the weighted minimum mean squared error (WMMSE) framework, establishes performance limits and motivates decentralized strategies. 
We subsequently introduce two decentralized optimization schemes, based on approximating the hardening bound and decentralizing the WMMSE framework, for two representative inter-satellite link (ISL) topologies, i.e., Ring and Star topologies. In the Ring topology-based beamforming scheme, satellites update beamformers locally and exchange intermediate parameters sequentially. On the other hand, in the Star topology-based beamforming scheme, edge satellites update beamformers locally and in parallel, achieving consensus on intermediate parameters at a central satellite using a penalty-dual decomposition (PDD) framework. 
Extensive simulations demonstrate that the proposed distributed beamforming schemes achieve similar performance with the centralized beamforming scheme while improving scalability significantly. Additionally, we reveal the delay–overhead trade-off between the two topologies.
\end{abstract}


\IEEEpeerreviewmaketitle
\section{Introduction}
5G is being deployed at an unprecedented pace, transforming daily life and driving innovation across global vertical industries. However, a significant digital divide persists. According to data released by the \ac{itu} in 2023, approximately one-third of the global population, about 2.6 billion people, remain offline~\cite{ITU2023DATA}. Bridging this gap remains a core ambition for 6G, with the vision to \emph{connect the unconnected}~\cite{Rethink6G}. Early 6G initiatives, both academic and industrial~\cite{imt2030vision,6GtakeShape}, have already emerged, with ubiquitous connectivity identified as one of the six key dimensions of the ``6G wheel'' defined by the \ac{itu}. From a standardization standpoint, the 3rd Generation Partnership Project (3GPP) has launched a study item on 5G \ac{ntn}, aiming to integrate satellite systems into mobile broadband and machine-type communication scenarios~\cite{3gpp.38.811}. It is widely anticipated that \ac{leo} satellite networks, as a fundamental component of \ac{ntn}, will be essential to achieving ubiquitous coverage in 6G and beyond, a feat unachievable by terrestrial networks alone.

\ac{leo} satellite systems feature dense constellation deployment, enhanced signal strength, and lower latency due to their reduced orbital altitude. As a result, they have already surpassed traditional \ac{meo} and \ac{geo} satellite systems in terms of communication quality. However, the conventional paradigm, where each \ac{ut} is served by a single \ac{leo} satellite, faces inherent limitations in power budget and array aperture. This poses challenges in meeting the growing demand for high-throughput communications using a single \ac{leo} satellite~\cite{6GtakeShape}. Consequently, there is increasing interest in leveraging the cooperative capabilities of multiple \ac{leo} satellites, interconnected via \acp{isl}, to jointly serve \acp{ut}~\cite{halim2022oj,halim2023oj,Bacci2023taes,kexin2024twc,poor2024tsp,moewin2025jsac,zack2025pace,meixia2024twc,Vanelli2024taes,Gaojie2024snt}. For examples, the authors of \cite{halim2022oj,halim2023oj} introduce the distributed massive \ac{mimo} concept, which allows \acp{ut} to be served by networked \ac{leo} satellites. Their results show that cooperation forms a virtually enlarged antenna array, thereby enabling distributed beamforming and enhancing communication rates. In \cite{Bacci2023taes}, the impact of the geometrical formation of these distributed arrays, hosted by networked satellites, on downlink communication throughput is thoroughly analyzed. In \cite{kexin2024twc} and \cite{poor2024tsp}, the authors exploit the antenna arrays at each \ac{ut} to spatially filter signals from multiple \ac{leo} satellites and to align them in both time and frequency domains. The optimality of single-stream transmission from each satellite to each \ac{ut} for maximizing the ergodic sum rate of networked \ac{leo} satellite distributed beamforming is validated in \cite{kexin2024twc}, while \cite{poor2024tsp} devises a hardware-friendly analog beamforming scheme. The works in \cite{moewin2025jsac,zack2025pace} leverage the \ac{los}-dominant nature of the \ac{leo} satellite channel and use \ac{ut} position information to facilitate efficient distributed beamforming across networked \ac{leo} satellites. In \cite{meixia2024twc}, a joint hybrid beamforming and user-scheduling scheme is designed for cooperative satellite networks.

Despite recent advancements in networked \ac{leo} satellite beamforming, two major limitations persist in conventional approaches. First, many existing works rely on instantaneous \ac{csi} for beamforming\cite{halim2022oj,halim2021vtm,meixia2024twc}, which may be difficult to obtain due to the latency incurred by the channel estimation process and the short coherence time resulting from the rapid movement of \ac{leo} satellites\cite{chang2023iotj,semiblind2025cl,blockKF2023cl,ming2025twc}. Second, many approaches adopt a centralized optimization paradigm, where each satellite uploads its \ac{csi} to a \ac{cpu}, typically implemented on a master or central satellite, that jointly computes the beamformers for all satellites and then distributes them back to the corresponding nodes\cite{moewin2023jsac,meixia2024twc,poor2024tsp}. 
Given the limited on-board processing capabilities of \ac{leo} satellites, this paradigm poses significant scalability challenges, as the central satellite must handle the majority of the computational burden.
To address these limitations, several distributed beamforming schemes have been proposed. For simplicity, they typically employ simple linear beamformers, such as \ac{mrt}, \ac{zf}, or their variants\cite{moewin2023jsac,meixia2024twc,halim2022oj,halim2023oj,Bacci2023taes,Vanelli2024taes}. While computationally attractive, these heuristic methods may lead to noticeable performance degradation compared to optimization-based methods that more effectively refine the solution toward the desired utility.

Therefore, a critical research gap exists in designing distributed beamforming schemes tailored for \ac{leo} satellite networks that do not rely on instantaneous \ac{csi}, yet deliver high performance and are scalable for practical deployment\footnote{While our focus is on algorithmic scalability under statistical \ac{csi}, practical transceiver/aperture hardware impairments (e.g., phase noise, nonlinearity, mutual coupling) can influence large-array \ac{ntn}/\ac{leo} beamforming\cite{qingchao2024RHS-HI,qingchao2024RHS-NF,qingchao2024SIM-CF,qingchao2025HM-LEO}. Incorporating such effects into the proposed framework is an important extension left for future work.}. To address this gap, this paper investigates distributed downlink beamforming schemes for scalable networked \ac{leo} satellites, aiming to unleash the power of large-scale cooperation over \ac{leo} satellite networks. Our goal is not merely to \emph{connect the unconnected}, but to \emph{uplift the connected} with transformative communication quality enhancements in the sky as we advance toward 6G. The main contributions of this paper are summarized as follows.

\begin{itemize}
\item We develop a comprehensive system model for networked \ac{leo} satellite distributed beamforming, where \ac{ofdm} is employed, and each satellite is equipped with a large antenna array and a limited number of \acp{rfc}. The model encompasses a thorough analysis of channel characteristics and pragmatic designs for user scheduling and analog beamforming for networked \ac{leo} satellite system. To sidestep the need for instantaneous \ac{csi}, we resort to the ergodic rate, which is a function of channel statistics, and then derive its closed-form lower bound by invoking the hardening bound. Using the derived ergodic rate expression, we formulate a downlink sum rate maximization problem that optimizes digital beamformers under the power budget constraints of each satellite. We first solve this problem using the \ac{wmmse} framework in a centralized manner, which not only lays the foundation for developing distributed optimization schemes but also establishes an upper bound for their performance;
\item We develop scalable distributed optimization approaches tailored to networked \ac{leo} satellite beamforming by approximating the hardening bound and modifying the \ac{wmmse} framework to enable decentralized processing.\footnote{Note that distributed implementations of WMMSE have been studied in prior work, e.g., \cite{shi2011wmmse}. However, those approaches rely on \ac{ut}-side covariance estimation and iterative feedback to the transmitters, which is infeasible in \ac{leo} satellite systems due to long propagation distances, limited feedback capacity, and excessive latency. In contrast, our design leverages \acp{isl} to achieve decentralized coordination without \ac{ut}-side involvement.} In the proposed design, beamformers are optimized locally at the corresponding satellites, while the exchange of intermediate parameters, crucial for achieving coherent utility, follows the information flow dictated by different yet representative \ac{isl} topologies, namely the Ring and Star topologies. In the Ring topology, information flows unidirectionally through the network in a sequential manner, which is fully decentralized but may incur longer delays due to propagation time. In the Star topology, bidirectional information exchange occurs between edge satellites and a central satellite node, enabling parallel processing at the edges. This Star topology, however, requires a central node to reach consensus on the intermediate parameters updated simultaneously by the edge nodes, a challenge that is addressed using a \ac{pdd}-based consensus algorithm.
\item We conduct extensive simulations on realistic networked \ac{leo} satellite systems to validate the effectiveness of the proposed approaches: 1) comparable performance is achieved relative to the centralized counterpart, while preserving scalability; 2) significant performance gains are achieved over conventional \ac{leo} satellite distributed beamforming schemes that use heuristic local beamformers, as well as schemes where each \ac{ut} is served by a single \ac{leo} satellite, either with optimized or heuristic beamforming; and 3) a tradeoff between delay and communication overhead is observed across different topologies, offering valuable insights for future research on large-scale cooperation in more complex and dynamic LEO satellite networks.
\end{itemize}

This paper is organized as follows. We begin by presenting the system model in Section \ref{sec-2}. Section \ref{sec-3} proposes a centralized optimization scheme for networked \ac{leo} satellite distributed beamforming that maximizes the downlink sum rate. Sections \ref{sec-4} and \ref{sec-5} introduce distributed optimization schemes, based on the Ring and Star topologies of \acp{isl}, respectively, to address the scalability issues of the centralized counterpart. Numerical results are presented in Section \ref{sec-6}, and Section \ref{sec-7} concludes the paper.

\emph{Notations:} Scalars are written in regular lowercase, while vectors and matrices are denoted by bold lowercase and bold uppercase letters, respectively. For a vector $\mathbf{a}$, its 2-norm is expressed as $\left\|\mathbf{a}\right\|$. For a matrix $\mathbf{A}$, its Frobenius norm is expressed as $\left\|\mathbf{A}\right\|_{\text{F}}$. We use $(\cdot)^*$, $(\cdot)^{\mathsf{T}}$ and $(\cdot)^{\mathsf{H}}$ as superscripts to indicate the conjugate, transpose, and Hermitian transpose operations, respectively. $\Re \left\{a\right\}$ denotes the real part of a complex number $a$. The Kronecker and Hadamard products are denoted by $\otimes$ and $\odot$, respectively. $\mathbb{E}[\cdot]$, $\mathbb{V}\left[\cdot\right]$, and $\text{vec}(\cdot)$ denote the expectation, variance, and vectorization operators, respectively. For statistical distributions, $\mathcal{CN}(\boldsymbol{\mu}, \mathbf{C})$ represents a circularly symmetric complex Gaussian distribution characterized by mean $\boldsymbol{\mu}$ and covariance matrix $\mathbf{C}$. $\mathbf{1}_N$ and $\mathbf{0}_N$ denote the all-one and all-zero $N$-dimensional column vectors, respectively. The notation $\mathbf{a} \ge \mathbf{b}$ denotes the inequality between each corresponding element of $\mathbf{a}$ and $\mathbf{b}$.

Tables \ref{abbre-add} and \ref{notation_table} detail the main abbreviations and essential symbols used throughout this paper, respectively.

\begin{table}[!htbp]
    \centering
    \caption{List of Abbreviations}
    \label{abbre-add}
    \begin{tabular}{@{}ll@{}}
        \toprule
        \textbf{Abbreviation} & \textbf{Full Form} \\ 
        \midrule
        AOD & Angle Of Departure \\
        ALP & Augmented Lagrangian Problem \\
        CPU & Central Processing Unit \\
        CSI & Channel State Information \\
        DOF & Degree Of Freedom \\
        GEO & Geostationary Earth Orbit \\
        GNSS & Global Navigation Satellite System \\
        LEO & Low Earth Orbit \\
        LOS & Line-Of-Sight \\
        MEO & Medium Earth Orbit \\
        MIMO & Multiple-Input-Multiple-Output \\
        MRT & Maximum Ratio Transmission \\
        NLOS & Non-Line-Of-Sight \\
        NTN & Non-Terrestrial Networks \\
        OFDM & Orthogonal Frequency Division Multiplexing \\
        PDD & Penalty Dual Decomposition \\
        QCQP & Quadratically Constrained Quadratic Program \\
        RFC & Radio Frequency Chain \\
        UPA & Uniform Planar Array \\
        UT & User Terminal \\
        WMMSE & Weighted Minimal Mean Squared Error \\
        ZF & Zero-Forcing \\
        \bottomrule
    \end{tabular}
\end{table}

\begin{table}[t]
    \centering
    \caption{List of Essential Symbols}
    \label{notation_table}
    \begin{tabular}{@{}ll@{}}
        \toprule
        \textbf{Notation} & \textbf{Definition} \\ 
        \midrule
        $S$ & LEO satellite number \\
        $U$ & UT number \\
        $N = N_{\text{h}} \times N_{\text{v}}$ & Antenna number (horizontal/vertical) \\
        $N_{\mathrm{RF}}$ & RFC number (per satellite) \\
        $T=\min(U,N_{\mathrm{RF}})$ & Served UT number (per satellite) \\
        $B$ & System (operating) bandwidth \\
        $\Delta f$ & Subcarrier spacing \\
        $K$ & Subcarrier number \\
        $T_{\mathrm{sym}}$ & OFDM symbol duration \\
        $\boldsymbol{\theta}_{s,u}=[\theta^{\mathrm{az}}_{s,u},\,\theta^{\mathrm{el}}_{s,u}]^{\mathsf{T}}$ & AOD (azimuth/elevation) \\
        $G(\theta^{\mathrm{el}}_{s,u})$ & Antenna radiation gain \\
        $\mathbf{a}(\boldsymbol{\theta}_{s,u})$ & Array steering vector \\
        $\alpha_{s,u}[k]$ & Composite channel gain (Rician) \\
        $\bar{\alpha}_{s,u}[k]$ & Mean of $\alpha_{s,u}[k]$ \\
        $\beta_{s,u}[k]$ & Variance of $\alpha_{s,u}[k]$ \\
        $\mathbf{F}_{s}=[\mathbf{f}_{s,1},\ldots,\mathbf{f}_{s,N_{\mathrm{RF}}}]$ & Analog beamformer \\
        $\mathbf{W}_{s}[k]=[\mathbf{w}_{s,1}[k],\ldots,\mathbf{w}_{s,U}[k]]$ & Digital beamformer (per subcarrier) \\
        $\boldsymbol{\delta}_{s}=[\delta_{s,1},\ldots,\delta_{s,U}]^{\mathsf{T}}$ & Scheduler vector \\
        \bottomrule
    \end{tabular}
\end{table}

\begin{figure}[t] 
		\centering
		\includegraphics[width=1.0 \linewidth]{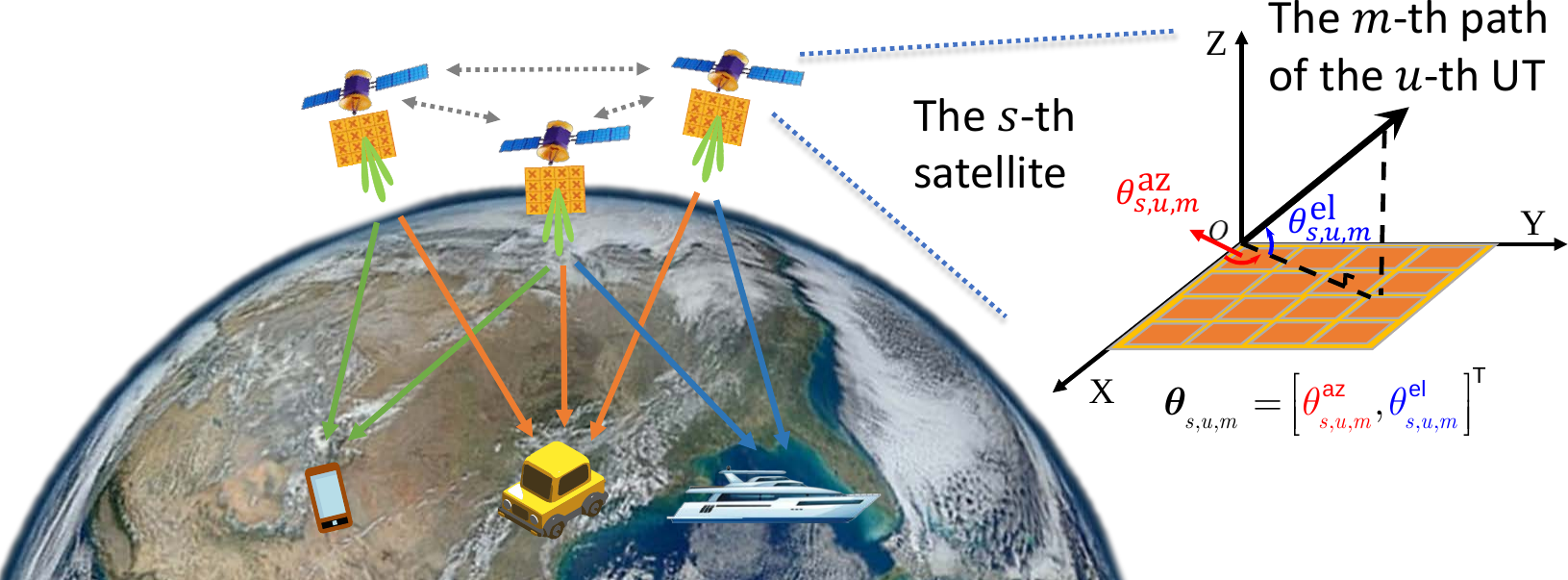}
		\caption{An illustration of networked-LEO satellite system, where multiple LEO satellites cooperatively serve UTs through distributed beamforming. The geometric configuration of the local coordinate systems for LEO satellites is depicted on the right part, where the UPA is deployed in the XY-plane.}
		\label{sys_mod}
\end{figure}

\section{System Model}\label{sec-2}
As illustrated in Fig. \ref{sys_mod}, we consider a networked-LEO satellite system where $S$ \ac{leo} satellites serve $U$ \acp{ut} in the downlink. Each \ac{ut} is equipped with a single antenna. Each satellite is equipped with a \ac{upa} comprising $N = N_{\text{h}} \times N_{\text{v}}$ half-wavelength-spaced antennas, where $N_{\text{h}}$ and $N_{\text{v}}$ denote the number of antennas in the horizontal and vertical dimensions, respectively. In addition, each satellite is equipped with $N_{\text{RF}}\le N$ \acp{rfc}. In our work, we assume an identical number of antennas and \acp{rfc} per satellite for simplicity. Nonetheless, it should be noted that extending the model to scenarios with varying antenna and/or \ac{rfc} configurations is straightforward. Each satellite serves $T = \min(U, N_{\text{RF}})$ \acp{ut}, within overlapping time-frequency resources according to a certain scheduling policy. Meanwhile, by exploiting \acp{isl}, information concerning different \acp{ut} is exchanged within the network, enabling each \ac{ut} to be served by multiple satellites simultaneously.

\subsection{Channel Model}
\subsubsection{Downlink Satellite Channel}
Let $B$ denote the operating bandwidth of the satellite system. We consider an \ac{ofdm} system with $K$ subcarriers, and carrier frequency being $f_c$, where $\Delta f = B/K$ and $T_{\text{sym}} = 1/\Delta f$ represent the subcarrier spacing and symbol duration, respectively. The downlink channel from the $s$-th satellite to the $u$-th \ac{ut} during the $\ell$-th \ac{ofdm} symbol on the $k$-th subcarrier is modeled as
\begin{align}\label{chan_mod} 
\mathbf{h}_{s,u}\left[\ell,k \right] = &\sum_{m=0}^{M_{s,u}}\alpha_{s,u,m} G\left(\theta_{s,u,m}^{\mathrm{el}}\right)e^{\jmath 2 \pi\left(\ell T_{\text{sym}} \upsilon_{s,u,m} - k \Delta f \tau_{s,u,m} \right) } \notag \\
&\times \mathbf{a}\left(\boldsymbol{\theta}_{s,u,m}\right),    
\end{align}  
where $M_{s,u}$ denotes the number of propagation paths, $\alpha_{s,u,m}$ is the complex channel gain. Here, the \ac{los} path is indexed by $m=0$, while the remaining paths are \ac{nlos}. The parameters $\tau_{s,u,m}$ and $\upsilon_{s,u,m}$ represent the propagation delay and Doppler shift, respectively, and $\mathbf{a}(\boldsymbol{\theta}_{s,u,m}) \in \mathbb{C}^{N}$ is the steering vector at the satellite, where $\boldsymbol{\theta}_{s,u,m} = [\theta_{s,u,m}^{\mathrm{az}},\theta_{s,u,m}^{\mathrm{el}}]^{\mathsf{T}}$ denotes the \ac{aod}, comprising both azimuth and elevation angles. Additionally, $G(\theta_{s,u,m}^{\mathrm{el}})$ denotes the antenna radiation pattern, which depends solely on the elevation angle \cite{balanis2005antenna}. Without loss of generality, we assume that the \ac{upa} at each satellite is deployed on the XY-plane of its local coordinate system, as illustrated in Fig.~\ref{sys_mod}. Let $\mathbf{n}(N) = [0,\ldots,N-1]^{\mathsf{T}}$. The steering vector is given by
\begin{equation}  
\mathbf{a}\left(\boldsymbol{\theta}_{s,u,m}\right) = e^{-\jmath 2 \pi \phi_{s,u,m}^{\text{h}} \mathbf{n}\left(N_{\text{h}}\right)} \otimes e^{-\jmath 2 \pi \phi_{s,u,m}^{\text{v}} \mathbf{n}\left(N_{\text{v}}\right)},  
\end{equation}  
where $\phi_{s,u,m}^{\text{h}} = d \cos \theta_{s,u,m}^{\mathrm{az}} \cos \theta_{s,u,m}^{\mathrm{el}}/\lambda$ and $\phi_{s,u,m}^{\text{v}} = d \sin \theta_{s,u,m}^{\mathrm{az}} \cos \theta_{s,u,m}^{\mathrm{el}}/\lambda$. Here, $d$ denotes the antenna spacing along each dimension, and $\lambda$ is the wavelength corresponding to the central carrier frequency $f_c$.


\subsubsection{Equivalent LOS Channel}
In \ac{leo} satellite systems, since the satellite altitude is significantly greater than the distances between the scatterers (typically located on the ground and around the receiver) and the \ac{ut}, the \acp{aod} in \eqref{chan_mod} can be approximated as identical across all paths, i.e., $\boldsymbol{\theta}_{s,u} = \boldsymbol{\theta}_{s,u,m},\forall m$. Note also that the Doppler shift associated with each path is induced by the movement of both the satellite and the \ac{ut}, such that $\upsilon_{s,u,m} = \upsilon_{s,u,m}^{\text{Sat}} + \upsilon_{s,u,m}^{\text{UT}}$, where $\upsilon_{s,u,m}^{\text{Sat}}$ and $\upsilon_{s,u,m}^{\text{UT}}$ denote the contributions to the Doppler shift from the satellite and the \ac{ut}, respectively. For the same reason as with the \acp{aod}, the satellite-induced Doppler shift can be considered identical across all paths, i.e., $\upsilon_{s,u}^{\text{Sat}} = \upsilon_{s,u,m}^{\text{Sat}},\forall m$. Let $\tau_{s,u} = \tau_{s,u,0}$ denote the minimum propagation delay corresponding to the \ac{los} path, and let $\tau_{s,u,m}^{\text{Diff}} = \tau_{s,u,m} - \tau_{s,u}$ represent the differential delay of the $m$-th path relative to the \ac{los} path. Then, we can rewrite \eqref{chan_mod} as the equivalent \ac{los} channel as
\begin{equation}\label{chan_mod_approx} 
\mathbf{h}_{s,u}\left[\ell,k \right] = \alpha_{s,u} \left[k \right]e^{\jmath 2 \pi\left(\ell T_{\text{sym}} \upsilon_{s,u}^{\text{Sat}} - k \Delta f \tau_{s,u} \right) }\Grave{\mathbf{a}}\left(\boldsymbol{\theta}_{s,u}\right),
\end{equation}
where $\Grave{\mathbf{a}}(\boldsymbol{\theta}_{s,u}) = G(\theta_{s,u}^{\mathrm{el}})\mathbf{a}(\boldsymbol{\theta}_{s,u})$ and $\alpha_{s,u}[k] = \sum_{m=0}^{M_{s,u}}\alpha_{s,u,m}e^{\jmath 2 \pi\left(\ell T_{\text{sym}} \upsilon_{s,u,m}^{\text{UT}} - k \Delta f \tau_{s,u,m}^{\text{Diff}} \right) }$ denotes the composite channel gain, which is modeled as a Rician random variable with Rician factor $\kappa_{s,u}$ and average power $\mathbb{E}[|\alpha_{s,u}[k]|^2] = \gamma_{s,u}[k]$ \cite{poor2024tsp,you2020jsac}. 

It is worth noting that $\alpha_{s,u}[k]$ is affected by the Doppler shift induced by \ac{ut} movement, and is therefore time-varying. However, since \ac{ut} movement is moderate in most practical scenarios, we henceforth assume that $\alpha_{s,u}[k]$ is slowly varying and omit its dependence on $\ell$. Specifically, the real and imaginary parts of $\alpha_{s,u}[k]$ are independent and identically distributed real-valued Gaussian random variables, each with the mean and variance given by 
\begin{equation}
\bar{\alpha}_{s,u}[k] = \sqrt{\frac{\kappa_{s,u}\gamma_{s,u}[k]}{2(1+\kappa_{s,u})}},\quad  \beta_{s,u}[k] = \frac{\gamma_{s,u}[k]}{2(1+\kappa_{s,u})}.   
\end{equation}

The large-scale path loss on the $k$-th subcarrier, i.e., $\gamma_{s,u}[k]$, can be represented in the dB scale as\cite{3gpp.38.811,poor2024tsp,meixia2024twc}
\begin{align}\label{large_loss_component}
-10\log_{10}\gamma_{s,u}[k] =\ & \text{PL}^{\text{FS}}_{s,u} + \text{PL}^{\text{SF}}_{s,u} + \text{PL}^{\text{CL}}_{s,u} + \text{PL}^{\text{AB}}_{s,u} \notag \\
&+ \text{PL}^{\text{SC}}_{s,u},
\end{align}
where $\text{PL}^{\text{FS}}_{s,u} = 20\log_{10}d_{s,u} + 20\log_{10}(f_c + k\Delta f) - 147.55$ denotes the free-space path loss \cite{3gpp.38.811}, with $d_{s,u}$ representing the distance from the $s$-th \ac{leo} satellite to the $u$-th \ac{ut}. The term $\text{PL}^{\text{SF}}_{s,u}$ captures the shadow fading component, while $\text{PL}^{\text{CL}}_{s,u}$ accounts for the clutter loss. The term $\text{PL}^{\text{AB}}_{s,u}$ models the atmospheric absorption loss, and $\text{PL}^{\text{SC}}_{s,u}$ represents the scintillation-induced attenuation caused by ionospheric and tropospheric irregularities. 

Several additional elaborations regarding the above channel model are in order:
\begin{itemize}
    \item In \ac{los}-dominated scenarios, both shadow fading and clutter loss typically have a negligible impact \cite{pathloss2021cl}. Accordingly, we omit these terms from the performance evaluation in our studied scenarios;
    \item The atmospheric absorption, which depends on signal frequency and satellite elevation angle, is calculated according to \ac{itu} recommendations \cite{ITU676_12} and configured using an off-the-shelf MATLAB script\cite{hourani2024atmospheric};
    \item Per \cite{3gpp.38.811}, ionospheric scintillation effects can be ignored for carrier frequencies beyond $6\ \text{GHz}$. As for the tropospheric scintillation term within $\text{PL}^{\text{SC}}_{s,u}$, which lacks a simple analytical form, we adopt the empirical model from \cite{pinjun2024jsac} as a reference.
\end{itemize}

\begin{remark}
From \eqref{chan_mod_approx}, the satellite–\ac{ut} link includes a phase term due to Doppler shift and propagation delay. In this work, we assume satellite-side, stream-level pre-compensation (timing advance and carrier pre-rotation) toward each \ac{ut} under a common time/frequency reference (e.g., via \ac{gnss}), so that signals from cooperating satellites arrive delay- and Doppler-aligned at the receiver, avoiding complicated \ac{ut}-side post-compensation. This idealized compensation is commonly adopted in the \ac{leo}-\ac{ntn} literature (e.g., \cite{you2020jsac,kexin2024twc,poor2024tsp,moewin2025jsac}), where predictable orbital dynamics and \ac{ut} positioning enable precise correction. Accordingly, \eqref{chan_mod_approx} simplifies to
\begin{equation}
\mathbf{h}_{s,u}\left[k \right] = \alpha_{s,u} \left[k\right]\Grave{\mathbf{a}}\left(\boldsymbol{\theta}_{s,u}\right).
\end{equation}
In practice, small orbital perturbations and oscillator imperfections lead to residual Doppler/delay errors. Within our statistical formulation, these residuals primarily manifest as additional random phase fluctuations in the composite gain $\alpha_{s,u}[k]$. Because our performance metric is the ergodic rate characterized via the hardening bound (introduced later), this metric is typically robust to such instantaneous phase drifts, which slightly alters its second-order statistics. A more detailed modeling and mitigation of imperfect compensation are beyond the scope of this paper and will be investigated in future work.
\end{remark}

\subsection{Signal Model}
As introduced, the number of \acp{rfc} employed by each \ac{leo} satellite can be (significantly) smaller than the number of antennas. This is a rational consideration for reducing hardware and energy costs, as each \ac{rfc} typically incurs high expense and power consumption, which are particularly critical in \ac{leo} satellite scenarios due to limited onboard resources and power supply \cite{meixia2024twc}. Consequently, we consider a hybrid beamforming architecture, where the \acp{rfc} are connected to the antennas through a cost-effective analog network implemented using \acp{ps}, enabling a hardware- and energy-efficient beamforming solution. 

Let $\mathbf{s}[\ell,k]= [s_1[\ell,k],\ldots,s_U[\ell,k]]^{\mathsf{T}}\sim \mathcal{CN}(\mathbf{0}_U,\mathbf{I}_U)$ denote the collection of data streams for $U$ \acp{ut} during the $\ell$-th symbol of the $k$-th subcarrier. The corresponding transmit signal at the $s$-th \ac{leo} satellite is expressed as
\begin{equation}
\mathbf{x}_s\left[\ell,k\right] = \mathbf{F}_s \mathbf{W}_s\left[k\right]\text{diag}\left(\boldsymbol{\delta}_s\right)\mathbf{s}\left[\ell,k\right],
\end{equation}
where $\mathbf{F}_s = [\mathbf{f}_{s,1},\ldots,\mathbf{f}_{s,N_{\text{RF}}}]\in \mathbb{C}^{N \times N_{\text{RF}}}$ denotes the analog beamformer, $\mathbf{W}_s\left[k\right] = [\mathbf{w}_{s,1}\left[k\right],\ldots,\mathbf{w}_{s,U}\left[k\right]] \in \mathbb{C}^{N_{\text{RF}} \times U}$ represents the digital beamformer, and $\boldsymbol{\delta}_s = [\delta_{s,1},\ldots,\delta_{s,U}]^{\mathsf{T}}$ denotes the scheduler, with each element being either $0$ or $1$. Specifically, the $u$-th \ac{ut} data stream is transmitted via the $s$-th satellite if $\delta_{s,u} = 1$, and is not transmitted if $\delta_{s,u} = 0$.

The signal received at the $u$-th \ac{ut} during the $\ell$-th symbol of the $k$-th subcarrier is given by  
\begin{align}\label{receive_sig}
y_u\left[\ell,k\right] &= \sum_{s=1}^{S} \mathbf{h}_{s,u}^{\mathsf{T}} \left[k\right]  \mathbf{F}_s \mathbf{W}_s\left[k\right]\text{diag}\left(\boldsymbol{\delta}_s\right)\mathbf{s}\left[\ell,k\right] + z_u\left[\ell,k\right] \notag \\
&= \sum_{s=1}^{S} \alpha_{s,u} \left[k\right]\mathbf{g}_{s,u}^{\mathsf{T}} \tilde{\mathbf{w}}_{s,u}\left[k\right] s_u\left[\ell,k\right]  \\
&\quad + \underbrace{\sum_{l\neq u}^{U} \sum_{s=1}^{S} \alpha_{s,u} \left[k\right]\mathbf{g}_{s,u}^{\mathsf{T}} \tilde{\mathbf{w}}_{s,l}\left[k\right] s_l\left[\ell,k\right]}_{\text{Inter-user interference (IUI)}} + z_u\left[\ell,k\right],\notag
\end{align}  
where $\mathbf{g}_{s,u} = \mathbf{F}_s^{\mathsf{T}}\Grave{\mathbf{a}}(\boldsymbol{\theta}_{s,u}) \in \mathbb{C}^{N_{\text{RF}}}$, $\tilde{\mathbf{w}}_{s,u}[k] = \delta_{s,u}\mathbf{w}_{s,u}[k]$, and $z_u[\ell,k]\sim \mathcal{CN}(\mathbf{0},\sigma^2)$ denotes the \ac{awgn}. Here, the noise variance is given by $\sigma^2 = N_0 \Delta f$, where $N_0$ is the single-sided \ac{psd}.

\subsection{User Scheduling}
In our work, instead of jointly optimizing the schedulers and beamformers, which typically involves a time-consuming iterative process and incurs significant computational complexity \cite{moewin2023jsac,meixia2024twc,dong2023twc,lei2024jsac}, we adopt a more pragmatic scheduling method based on geometric relationships between satellites and \acp{ut}. Let $\mathcal{S} = \{1,2,\ldots,S\}$ and $\mathcal{U} = \{1,2,\ldots,U\}$ denote the index sets of \ac{leo} satellites and \acp{ut}, respectively. For each $s \in \mathcal{S}$, let $\mathcal{U}_s = \{u_s[1],\ldots,u_s[T]\}$ be the index set (in \emph{ascending} order) of \acp{ut} served by the $s$-th satellite. The elements of $\mathcal{U}_s$ corresponds to the $T$ nearest \acp{ut} to the $s$-th satellite, which can be straightforwardly determined by sorting the distances from the \acp{ut} to the satellite.

Several points are worth emphasizing: \emph{First}, this scheduling method is independent of \ac{csi} and relies solely on \emph{geometric information}, such as the positions of the \acp{ut}, which vary slowly (compared to \ac{csi}) and can be obtained by the \acp{ut} via \ac{gnss} and fed back to the \ac{leo} satellites\footnote{Position feedback may be slightly outdated due to propagation/feedback latency. At \ac{leo} ranges (on the order of 500 km), one-way propagation is $<2$ ms. Even at 20 m/s, a \ac{ut} moves $<0.04$ m in this interval, negligible relative to the link distance, so geometry-dependent statistics (and thus statistical \ac{csi}) are essentially unaffected. Consequently, modest update rates (sub-second to multi-second, depending on mobility) typically suffice.}; \emph{Second}, beyond its practicality, a key rationale behind this scheduling method is that, in the \ac{los}-dominant satellite communication environment, the link budget is primarily determined by the communication distance. As a result, each satellite can efficiently utilize its resources and enhance the overall system performance; \emph{Third}, when $N_{\text{RF}} \ge U$, i.e., $T = U$, all \acp{ut} in $\mathcal{U}$ can be served by each satellite without requiring additional distance-based scheduling. However, since the number of \acp{ut} is typically much larger than that of \acp{rfc}, this scenario is not particularly relevant in practice.

\begin{remark}
It should be pointed out that, in dense deployments where multiple \acp{ut} are geographically close and thus exhibit nearly identical distances and similar large-scale/channel statistics, nearby \acp{ut} are aggregated into a group and the group is treated as a single scheduled entity. A single data stream is then transmitted to the entire group via beam-wise broadcasting (multicast). The distance-based scheduler operates on the group centroids and selects the $T$ nearest groups per satellite, which naturally avoids tie cases among nearly co-located \acp{ut}. Crucially, this grouping is agnostic to the subsequent beamforming optimization and a group is indistinguishable from an individual \ac{ut} after replacing the \ac{ut}'s geometry/statistics with those of the group's centroid (or aggregate statistics). The problem formulation, variables, constraints, and algorithms therefore remain unchanged.
\end{remark}

\newcounter{MYtempeqncn}
\begin{figure*}[!b]
\normalsize
\setcounter{MYtempeqncn}{\value{equation}}
\setcounter{equation}{10}
\hrulefill
\vspace{+5mm}
\begin{subequations}\label{rate_lb}
\begin{align}
R_u^{\text{LB}}\left[k\right] =& \log_2 \left(1 + \frac{\left|\mathbb{E}\left[\sum_{s=1}^{S} \alpha_{s,u} \left[k\right]\mathbf{g}_{s,u}^{\mathsf{T}} \tilde{\mathbf{w}}_{s,u}\left[k\right]\right]\right|^2}{\mathbb{V}\left[\sum_{s=1}^{S} \alpha_{s,u} \left[k\right]\mathbf{g}_{s,u}^{\mathsf{T}} \tilde{\mathbf{w}}_{s,u}\left[k\right]\right] + \sum_{l\ne u}^{U}\mathbb{E}\left[ \left| \sum_{s=1}^{S} \alpha_{s,u} \left[k\right]\mathbf{g}_{s,u}^{\mathsf{T}} \tilde{\mathbf{w}}_{s,l}\left[k\right]\right|^2 \right] + \sigma^2}\right)\label{rate_lb_a}\\
=&\log_2\left(1 + \frac{\left|\sum_{s=1}^{S} \bar{\alpha}_{s,u} \left[k\right]\mathbf{g}_{s,u}^{\mathsf{T}} \tilde{\mathbf{w}}_{s,u}\left[k\right] \right|^2}{\sum_{s=1}^{S}\beta_{s,u}\left[k\right] \left| \mathbf{g}_{s,u}^{\mathsf{T}} \left[k\right]\tilde{\mathbf{w}}_{s,u}\left[k\right]\right|^2 + \sum_{l\ne u}^{U} \tilde{\mathbf{w}}_u^{\mathsf{H}}\left[k\right] \mathbf{T}_u\left[k\right]  \tilde{\mathbf{w}}_u\left[k\right] + \sigma^2 }\right)\label{rate_lb_b}
\end{align}
\end{subequations}
\setcounter{equation}{\value{MYtempeqncn}}
\end{figure*}

\subsection{Analog Beamformer}\label{sec-ana-bf}
Inspired by \cite{liang2014wcl,nasir2020tvt}, the analog beamformer is designed using the steering vector as
\begin{equation}\label{analog_bf}
\mathbf{F}_s = \left[\mathbf{a}^*(\boldsymbol{\theta}_{s,u_s\left[1\right]}),\ldots,\mathbf{a}^*(\boldsymbol{\theta}_{s,u_s\left[T\right]})\right].
\end{equation}
Note that $\mathbf{F}_s$ is a function of the \acp{aod} $\boldsymbol{\theta}_{s,u}, u\in \mathcal{U}_s$, which are determined by the positions of \ac{leo} satellites and \acp{ut}. This enables the analog beamformers to be constructed \emph{locally} at each satellite and inherently satisfies the unit-modulus constraint of the \ac{ps} \cite{heath2016jstsp}. In addition, as future \ac{leo} satellite systems are envisioned to be equipped with massive antenna arrays \cite{you2020jsac,you2024twc,kexin2023twc,kexin2024twc}, this design allows us to harness the large \emph{array gain} provided by the excessive antennas in the analog domain \cite{liang2014wcl}, without relying on complex unit-modulus-constrained hybrid beamforming optimization. On the other hand, owing to the \ac{los}-dominant nature of the \ac{leo} satellite channel and the asymptotic orthogonality property (i.e., favorable propagation property)\cite{sun2015tcom}, the analog beamformer design in \eqref{analog_bf} enables us to concentrate the power of the equivalent steering vector $\mathbf{g}_{s,u}$ on its $u$-th element (if $u \in \mathcal{U}_s$), effectively managing IUI in the analog domain. This approach has been shown to achieve performance comparable to even fully digital solutions \cite{nasir2020tvt}.

\section{Centralized Beamforming Optimization and Inter-Satellite Link Topologies}\label{sec-3}
In this section, we first formulate a beamforming optimization problem for networked \ac{leo} satellites systems, with the objective of enhancing downlink sum rate. A \ac{wmmse}-based solution is then developed, which inherently requires centralized implementation at a \ac{cpu}, limiting its scalability in large-scale satellite networks. To address this issue, we introduce several distinct yet representative \ac{isl} topologies for \ac{leo} satellite constellations. These topologies serve as the foundation for developing distributed beamforming schemes based on decentralized optimization, thereby enabling the computational workload to be distributed among all participating \ac{leo} satellites.

\subsection{Sum Rate Maximization Problem Formulation}
Since acquiring accurate instantaneous \ac{csi} is challenging in \ac{leo} satellite systems due to short coherence time and long propagation delay (compared to terrestrial counterparts) \cite{kexin2023twc,chang2023iotj,semiblind2025cl,blockKF2023cl,ming2025twc}, we opt to optimize a statistical performance metric, typically characterized by the ergodic rate, instead of its instantaneous counterpart. Since the ergodic rate usually lacks a closed-form expression, we resort to its lower bound, namely the hardening bound \cite{Marzetta2016mMIMO,caire2018twc}, to facilitate tractability. 

From \eqref{receive_sig}, the lower bound of the ergodic rate at the $u$-th \ac{ut} over the $k$-th subcarrier is given in \eqref{rate_lb} at the bottom of this page.
Here, $\tilde{\mathbf{w}}_{u}[k] = [\tilde{\mathbf{w}}_{1,u}^{\mathsf{T}}[k],\ldots,\tilde{\mathbf{w}}_{S,u}^{\mathsf{T}}[k]]^{\mathsf{T}}$ and $\mathbf{T}_u[k] \in \mathbb{C}^{N_{\text{RF}} \times N_{\text{RF}}}$. For all $i,j \in \mathcal{S}$, we have
\begin{align}
\mathbf{T}_u^{i,j}\left[k\right]=\left\{ {\begin{array}{*{20}{l}}
\bar{\alpha}_{i,u}^*\left[k\right]\bar{\alpha}_{j,u}\left[k\right]\mathbf{g}_{i,u}^* \mathbf{g}_{j,u},&i \ne j,\\
\gamma_{i,u}[k]\mathbf{g}_{i,u}^*\mathbf{g}_{i,u},&i = j,\\
\end{array}} \right.    
\end{align}
where $\mathbf{T}_u^{i,j}[k] = [\mathbf{T}_u[k]]_{1 + (i-1)N_{\text{RF}}\,:\,iN_{\text{RF}},\,1 + (j-1)N_{\text{RF}}\,:\,jN_{\text{RF}}}$. Note that the transition from \eqref{rate_lb_a} to \eqref{rate_lb_b} is obtained through straightforward algebraic manipulation, and the detailed steps are omitted for brevity. From the expression in \eqref{rate_lb_b}, we observe that the communication rate now depends solely on \emph{statistical channel parameters} rather than instantaneous ones.

In the following, we formulate an optimization problem with respect to the digital beamformers (\emph{hereafter referred to simply as beamformers}) across multiple \ac{leo} satellites, i.e., $\mathbf{W}_s[k],\forall s$, aiming to maximize the sum rate across all \acp{ut}. To reduce algorithmic complexity, we assume that the transmit power is uniformly allocated across subcarriers, allowing the beamformers to be designed independently for each subcarrier. This assumption aligns with practical protocols, where beamforming in \ac{ofdm} systems is typically performed independently across different resource blocks~\cite{3gpp.38.802}. \emph{Accordingly, we omit the dependence on the subcarrier index $k$ for notational simplicity.} The sum rate maximization problem, under per-satellite power budget, is then formulated as
\addtocounter{equation}{1}
\begin{subequations}\label{ori_prob}
\begin{align}
\mathop {\max }\limits_{\tilde{\mathbf{W}}_s} \;\;\; &\sum_{u=1}^{U} R_u^{\text{LB}} \notag \\
{\rm{s.t.}}\;\;\;
& \left(1- \delta_{s,u}\right)\tilde{\mathbf{w}}_{s,u} = \mathbf{0}_{ N_{\text{RF}}},\;\forall s,u,\label{ori_prob_schedule}\\
&\left\| \mathbf{F}_s \tilde{\mathbf{W}}_s\right\|^2_{\text{F}}  \le \frac{P_s}{K}, \;\forall s,\label{ori_prob_pow_bud}
\end{align}  
\end{subequations}
where $\tilde{\mathbf{W}}_s = [\tilde{\mathbf{w}}_{s,1},\ldots,\tilde{\mathbf{w}}_{s,U}]$, and $P_s$ denotes the power budget of the $s$-th \ac{leo} satellite. Here, \eqref{ori_prob_schedule} ensures that $\tilde{\mathbf{w}}_{s,u} = \mathbf{0}_{ N_{\text{RF}}}$ whenever $\delta_{s,u} = 0$, i.e., the $u$-th \ac{ut} is not be served by the $s$-th satellite. 

\subsection{WMMSE-Based Beamformer}
In the following, we adopt the \ac{wmmse} optimization framework to reformulate the non-convex problem \eqref{ori_prob} into a sequence of convex quadratic problems by introducing auxiliary variables\cite{shi2011wmmse,zack2025pace}. This transformation enables efficient iterative optimization with guaranteed convergence to a stationary point\cite{shi2011wmmse}. Define the function 
\begin{align}\label{Upsilon_express}
\Upsilon_{u} = &\left|1 - \mu_{u}\sum_{s=1}^{S} \bar{\alpha}_{s,u} \mathbf{g}_{s,u}^{\mathsf{T}} \tilde{\mathbf{w}}_{s,u}\right|^2  \\
&+\left|\mu_{u}\right|^2\left(\underbrace{\sum_{s=1}^{S}\beta_{s,u} \left| \mathbf{g}_{s,u}^{\mathsf{T}} \tilde{\mathbf{w}}_{s,u}\right|^2 +
\sum_{l\ne u}^{U}\tilde{\mathbf{w}}_u^{\mathsf{T}} \mathbf{T}_u\tilde{\mathbf{w}}_u + \sigma^2}_{\Psi_u}\right). \notag   
\end{align}
Here, $\mu_{u}$ and $\nu_{u}$ serve as the auxiliary variables \cite{shi2011wmmse}. With this reformulation, problem \eqref{ori_prob} becomes equivalent to the following
\begin{align}\label{prob_refor}  
\mathop {\max }\limits_{\mu_{u},\nu_{u},\tilde{\mathbf{W}}_s} \;\;\; &\sum_{u=1}^{U} \left( \ln \nu_{u} - \nu_{u} \Upsilon_{u}\right) \notag \\  
\text{{\rm s.t.}} \;\;\;&  
\eqref{ori_prob_schedule},\eqref{ori_prob_pow_bud}.
\end{align}  
This problem can be efficiently solved through the alternating updates of $\tilde{\mathbf{W}}_s$, $\nu_{u}$, and $\mu_{u}$, as described below.

\subsubsection{Update of $\mu_{u}$}  
For fixed $\nu_{u}$ and $\tilde{\mathbf{W}}_s$, the optimal $\mu_{u}$ is obtained by minimizing $\Upsilon_{u}$ with respect to $\mu_{u}$, i.e., by setting $\partial \Upsilon_{u} / \partial \mu_{u} = 0$. This yields the closed-form solution
\begin{equation}\label{update_mu}  
\mu_{u} = \frac{\left(\sum_{s=1}^{S} \bar{\alpha}_{s,u} \mathbf{g}_{s,u}^{\mathsf{T}} \tilde{\mathbf{w}}_{s,u}\right)^{\mathsf{*}}}{\left|\sum_{s=1}^{S} \bar{\alpha}_{s,u} \mathbf{g}_{s,u}^{\mathsf{T}} \tilde{\mathbf{w}}_{s,u}\right|^2 + \Psi_u}.
\end{equation}

\subsubsection{Update of $\nu_{u}$}  
With $\mu_{u}$ and $\tilde{\mathbf{W}}_s$ fixed, the optimal $\nu_{u}$ that maximizes the objective in \eqref{prob_refor} is given by
\begin{equation}\label{update_nu}  
\nu_{u} = \frac{1}{\Upsilon_{u}}.  
\end{equation}

\subsubsection{Update of $\tilde{\mathbf{W}}_s$}\label{update_bf}  
Given the updated values of $\mu_{u}$ and $\nu_{u}$, the optimal beamforming matrix $\tilde{\mathbf{W}}_s$ is found by solving the following optimization problem
\begin{align}\label{cent_bf_opt}  
\mathop {\min }\limits_{\tilde{\mathbf{W}}_s} \;\;\; &\sum_{u=1}^{U} \nu_{u} \Upsilon_{u} \notag \\  
\text{{\rm s.t.}} \;\;\;  
&\eqref{ori_prob_schedule},\eqref{ori_prob_pow_bud}.  
\end{align}  
Here, $\Upsilon_{u}$ is convex quadratic function of $\tilde{\mathbf{W}}_s$. Therefore, \eqref{cent_bf_opt} constitutes a convex \ac{qcqp}, which can be efficiently solved using convex optimization solvers such as CVX.

\renewcommand{\algorithmicrequire}{\textbf{Input:}}
\renewcommand{\algorithmicensure}{\textbf{Output:}}
\begin{algorithm}[t]
\caption{Centralized Optimization for \ac{wmmse}-Based Networked \ac{leo} Satellite Distributed Beamforming}
\label{wmmse_cent}
\begin{algorithmic}[1]
\State \textbf{Initialize}: {$\tilde{\mathbf{W}}_s[k]$, $\forall s,k$;}
\For {$k = 1 : K$}
\Repeat
\State {Update $\mu_{u}[k]$ using \eqref{update_mu};}
\State {Update $\nu_{u}[k]$ using \eqref{update_nu};}
\State {Update $\tilde{\mathbf{W}}_s[k]$ by solving \eqref{cent_bf_opt} via CVX;}
\Until {the relative reduction in the objective value falls below a predefined threshold or a maximum number of iterations is reached;}
\EndFor
\State \textbf{Output}: {$\tilde{\mathbf{W}}_s[k],\forall s,k$.}
\end{algorithmic}
\end{algorithm}

\begin{figure}[t] 
		\centering
		\includegraphics[width=0.9\linewidth]{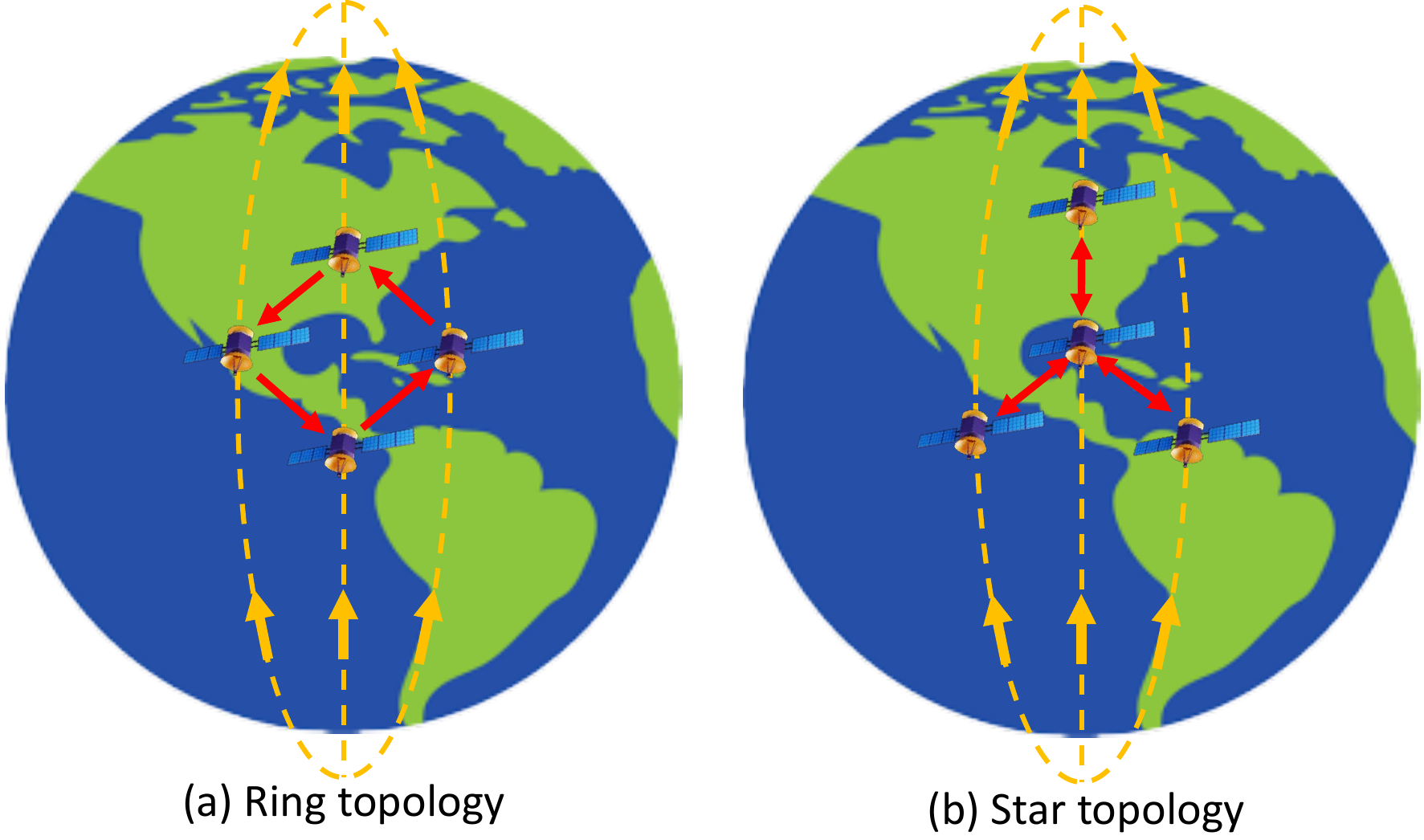}
		\caption{Illustrations of different \ac{isl} topologies: (a) Ring topology and (b) Star topology.}
		\label{constellation}
\end{figure} 

The iterative process for solving \eqref{ori_prob} is outlined in Algorithm \ref{wmmse_cent}. For completeness, we reintroduce the previously omitted subcarrier index $k$. The convergence of Algorithm \ref{wmmse_cent} is established in \cite{shi2011wmmse}. The computational complexity of each iteration is primarily determined by solving \eqref{cent_bf_opt}, which, as a \ac{qcqp} problem, is characterized by $\mathcal{O}((SN_{\text{RF}}T)^3)$.

\newcounter{MYtempeqncn2}
\begin{figure*}[!b]
\normalsize
\setcounter{MYtempeqncn}{\value{equation}}
\setcounter{equation}{17}
\hrulefill
\vspace{+5mm}
\begin{subequations}\label{rate_lb_new}
\begin{align}
R_u^{\text{LB}} =& \log_2 \left(1 + \frac{\left|\sum_{s=1}^{S} \bar{\alpha}_{s,u} \mathbf{g}_{s,u}^{\mathsf{T}} \tilde{\mathbf{w}}_{s,u}\right|^2}{\sum_{s=1}^{S}\beta_{s,u} \left| \mathbf{g}_{s,u}^{\mathsf{T}} \tilde{\mathbf{w}}_{s,u}\right|^2 + \sum_{l\ne u}^{U}\sum_{s=1}^{S}\left( \gamma_{s,u} \left| \mathbf{g}_{s,u}^{\mathsf{T}} \tilde{\mathbf{w}}_{s,l}\right|^2 + \eta_{s,l}\right) + \sigma^2}\right)\label{rate_lb_new_a}\\
\approx& \log_2 \left(1 + \frac{\left|\sum_{s=1}^{S} \bar{\alpha}_{s,u} \mathbf{g}_{s,u}^{\mathsf{T}} \tilde{\mathbf{w}}_{s,u}\right|^2}{\sum_{s=1}^{S}\beta_{s,u} \left| \mathbf{g}_{s,u}^{\mathsf{T}} \tilde{\mathbf{w}}_{s,u}\right|^2 + \sum_{l\ne u}^{U}\sum_{s=1}^{S}\gamma_{s,u} \left| \mathbf{g}_{s,u}^{\mathsf{T}} \tilde{\mathbf{w}}_{s,l}\right|^2 + \sigma^2}\right)\label{rate_lb_new_b}
\end{align}
\end{subequations}
\setcounter{equation}{\value{MYtempeqncn}}
\end{figure*}

\begin{remark}\label{remark_scalability}
Several important observations regarding computational complexity and scalability are worth highlighting:
\begin{itemize}
    \item The analog beamformers are directly constructed via \eqref{analog_bf} without being included in the optimization loop. As a result, the computational complexity is significantly reduced from being dominated by the number of antennas $N$ to being governed by the number of \ac{rfc}s $N_{\text{RF}}$. This complexity reduction is particularly beneficial for future \ac{leo} satellite systems, which are expected to employ extremely massive antenna arrays\cite{you2020jsac}.    
    \item As indicated in \eqref{ori_prob}, the digital beamformers at the \ac{leo} satellites, i.e., the $\tilde{\mathbf{W}}_s, \forall s$, are coupled within the objective function. Consequently, the \ac{wmmse}-based optimization framework necessitates a centralized implementation. In practice, a designated \ac{cpu} among the \ac{leo} satellites collects the statistical \ac{csi} reported by all participating satellites and computes the intermediate parameters required for optimization, such as $\mathbf{v}_u$, $\mathbf{q}_u$, and $\mathbf{T}_u$. The beamformers are then jointly optimized at the \ac{cpu} and subsequently distributed to the respective satellites. However, this centralized strategy imposes a heavy computational load on the \ac{cpu}, leading to serious scalability concerns as the algorithmic complexity scales cubically with the number of satellites.
\end{itemize}
\end{remark}

Remark \ref{remark_scalability} motivates us to optimize distributed beamforming in a decentralized manner by offloading computational tasks among the involved \ac{leo} satellites through the exploitation of \acp{isl}, while \emph{ensuring coherent computation toward optimizing a shared utility function}, such as the sum rate in this case. To this end, we briefly introduce \ac{isl} topologies below, which give rise to different decentralized beamforming designs.

\subsection{ISL Topologies}
\acp{isl} serve as space infrastructure that enables satellites to communicate directly with each other via RF and/or free-space optical (FSO) signals, rather than relying on ground stations, which suffer from limited link budgets and high latency. \acp{isl} offer high-speed and real-time communication capabilities for exchange of command, control, and data information~\cite{halim2021vtm,popov2020access}. \acp{isl} can be categorized into intra-plane and inter-plane types. Intra-plane \acp{isl} allow satellites within the same orbital plane to communicate with each other. These links are relatively stable, as the satellites move at the same speed and maintain a consistent distance. In contrast, inter-plane \acp{isl} enable communication between satellites in different orbital planes. These links are more dynamic because the relative distance and visibility between different orbits can vary significantly over time.

As illustrated in Fig. \ref{constellation}, we present two representative topologies, namely, the Ring topology (Fig. \ref{constellation}(a)) and the Star topology (Fig. \ref{constellation}(b)), spanning across multiple orbital planes\cite{liz2022access}. In the Ring topology, information flows unidirectionally through the network. In contrast, the Star topology enables bidirectional information exchange between edge satellites and a central satellite node. We implicitly assume that cooperation exists among \ac{leo} satellites in adjacent orbital planes, which traverse in the same direction\cite{popov2020access}. This assumption facilitates the maintenance of stable \acp{isl} topologies among cooperative \ac{leo} satellites and helps avoid frequent re-establishment of \acp{isl} caused by changes in physical topology. In what follows, we develop two decentralized implementations of distributed beamforming tailored to networked \ac{leo} satellites systems, corresponding to the Ring and Star topologies, respectively\footnote{In this work, the \ac{isl} topology is modeled as \emph{piecewise constant} within each beamforming-optimization window, whose duration is sub-second (typically tens to a few hundreds of ms). By contrast, handovers for a terrestrial \ac{ut} at 500–600 km LEO altitudes occur on timescales of tens of seconds to a few minutes. This separation of timescales allows the distributed algorithms to converge under a fixed topology and only refresh when a handover or major topology change occurs. In practice, a handover triggers rescheduling with a warm start from the previous solution, which primarily increases convergence time rather than altering the optimizer or update rules.}. 

Note that practical \ac{isl} topologies often depart from the two idealized structures considered here and may involve more complex, time-varying configurations and constraints~\cite{liz2022access}. In addition, link-layer effects, such as synchronization jitter and queuing, introduce per-hop latency, and occasional packet losses may trigger retransmissions. These aspects primarily increase the elapsed time to convergence but do not fundamentally alter the optimization framework developed in the following sections. Thus, for simplicity, we abstract them away to focus on the algorithmic design. A comprehensive treatment, covering decentralized algorithms for \emph{arbitrary} \ac{isl} topologies and practical mitigations (e.g., time-stamping and discarding stale messages, or reusing last-known intermediates when an update is missed) is left to future work.

\section{Distributed Beamforming Optimization Over Ring Topology-Based ISLs}\label{sec-4}
In this section, we present a networked \ac{leo} satellite distributed beamforming scheme based on the Ring \ac{isl} topology. In this approach, the exchange of optimization intermediates among satellites is facilitated through a ring-structured data flow. We begin by revisiting \eqref{rate_lb}, i.e., the lower bound of the ergodic rate, and approximating it in a manner amenable to decentralized computation. Then, we explain the decentralized \ac{wmmse} framework built upon the Ring topology-based \acp{isl}.

\subsection{Revisit of Hardening Bound of Ergodic Rate}\label{sec-4-a}
The expression in \eqref{rate_lb_b}, though compact, is \ac{ut}-centric and thus not suitable for satellite-centric decentralization. By omitting the dependence of the hardening bound on the subcarrier index for brevity, we rewrite \eqref{rate_lb_a} as \eqref{rate_lb_new_a} at the bottom of this page, where $\eta_{s,l} = \sum_{j \ne s}^{S}(\bar{\alpha}_{s,u}\mathbf{g}_{s,u}^{\mathsf{T}} \tilde{\mathbf{w}}_{s,l})^* (\bar{\alpha}_{j,u} \mathbf{g}_{j,u}^{\mathsf{T}} \tilde{\mathbf{w}}_{j,l})$, $\forall l \in \mathcal{S}{/}\{u\}$. ote that $\eta_{s,l}$ represents an IUI component that depends on coupled beamformers across multiple satellites. As a result, the contribution from any individual satellite cannot be separated without accounting for the others. This coupling complicates decentralized processing, as will be discussed later. Meanwhile, satellite communication typically operates in a noise-dominant regime due to severe signal attenuation over long propagation distances. In addition, as shown in Section \ref{sec-ana-bf}, the analog beamformer in \eqref{analog_bf} inherently suppresses IUI. Leveraging this property, we neglect the $\eta_{s,l}$ term in $R_u^{\text{LB}}$ to derive the approximated ergodic rate in \eqref{rate_lb_new_b}.

\subsection{Decentralization of WMMSE Framework}
After substituting the approximation given in \eqref{rate_lb_new_b} into \eqref{ori_prob}, we follow a reformulation similar to that presented in \eqref{Upsilon_express}. The only difference is that the original expression of $\Upsilon_u$ in \eqref{Upsilon_express} is now replaced by
\addtocounter{equation}{1}
\begin{equation}\label{gamma_express}
\Upsilon_u = \left|1- \mu_u F_u \right|^2 + \left|\mu_u\right|^2\left(P_u + \sum_{l\ne u}^{U}Q_{u,l} + \sigma^2 \right),       
\end{equation}
where
\begin{subequations}\label{inter_var}
\begin{align}
F_u &= \bar{\alpha}_{s,u} \mathbf{g}_{s,u}^{\mathsf{T}} \tilde{\mathbf{w}}_{s,u} + \underbrace{\sum_{j\ne s}^{S} \bar{\alpha}_{j,u} \mathbf{g}_{j,u}^{\mathsf{T}} \tilde{\mathbf{w}}_{j,u}}_{F_{s,u}}, \\
P_u &= \beta_{s,u} \left| \mathbf{g}_{s,u}^{\mathsf{T}} \tilde{\mathbf{w}}_{s,u}\right|^2 + \underbrace{\sum_{j\ne s}^{S}\beta_{j,u} \left| \mathbf{g}_{j,u}^{\mathsf{T}} \tilde{\mathbf{w}}_{j,u}\right|^2}_{P_{s,u}},\label{inter_var_P}\\
Q_{u,l} &= \gamma_{s,u} \left| \mathbf{g}_{s,u}^{\mathsf{T}} \tilde{\mathbf{w}}_{s,l}\right|^2 + \underbrace{\sum_{j\ne s}^{S}\gamma_{j,u} \left| \mathbf{g}_{j,u}^{\mathsf{T}} \tilde{\mathbf{w}}_{j,l}\right|^2}_{Q_{s,u,l}}\label{inter_var_Q}.
\end{align}
\end{subequations}

We emphasize through \eqref{inter_var} that the intermediate parameters $F_u$, $P_u$, and $Q_{u,l}$, which aggregate contributions from all satellites, can be decomposed into a component solely contributed by the $s$-th satellite and another aggregating the remaining $S-1$ satellites, i.e., $F_{s,u}$, $P_{s,u}$, and $Q_{s,u,l}$. At the $s$-th satellite, $F_u$, $P_u$, and $Q_{u,l}$ are made available through \acp{isl}, we subsequently show that the auxiliary variables $\mu_u$ and $\nu_u$, as well as the corresponding beamformer $\tilde{\mathbf{W}}_s$, can be updated \emph{locally} (previously done at a \ac{cpu}).

\subsubsection{Local Update of $\mu_u$ and $\nu_u$}
Given $F_u$, $P_u$, and $Q_{u,l}$, by applying the update rules for auxiliary variables in the \ac{wmmse} framework, it is straightforward to derive the update equations for $\mu_u$ and $\nu_u$as
\begin{equation}
\mu_u = \frac{F_u^*}{\left|F_u\right|^2 + P_u + \sum_{l\ne u}^{U}Q_{u,l} + \sigma^2}, \quad
\nu_u = \frac{1}{\Upsilon_u}.   \label{update_mu_and_nu_new}   
\end{equation}

\subsubsection{Local Update of $\tilde{\mathbf{W}}_s$}
With the updated values of $\mu_{u}$ and $\nu_{u}$, similar to \eqref{cent_bf_opt}, the update of $\tilde{\mathbf{W}}_s$ should follow the principle of minimizing the objective function $\sum_{u=1}^{U}\nu_u \Upsilon_u$, which aggregates contributions from all satellites. To enable the local update of $\tilde{\mathbf{W}}_s$, we first reformulate the objective function (\emph{by dropping the irrelevant terms regarding $\tilde{\mathbf{W}}_s$}) into 
\begin{equation}\label{refor_obj}
\sum_{u=1}^{U} \nu_{u} \Upsilon_{u} \Rightarrow-2\sum_{s=1}^{S}\Omega_{s} +\sum_{u=1}^{U} \nu_u \left|\mu_u\right|^2 Z_u ,
\end{equation}
where $\Omega_s = \sum_{u=1}^{U}\nu_u \Re\{\mu_u\bar{\alpha}_{s,u} \mathbf{g}_{s,u}^{\mathsf{T}} \tilde{\mathbf{w}}_{s,u}\}$ and $Z_u = |F_u|^2 +  P_u + \sum_{l\ne u}^{U} Q_{u,l}$. Note that $\Omega_s$ denotes the component exclusively contributed by the $s$-th satellite. Here, the variables $F_{s,u}$, $P_{s,u}$, and $Q_{s,u,l}$, which are extracted from $F_u$, $P_u$, and $Q_{u,l}$, respectively, as defined in \eqref{inter_var}, are treated as \emph{frozen} within $Z_u$, so that only the beamformer associated with the $s$-th satellite, i.e., $\tilde{\mathbf{W}}_s$ remains involved in the optimization. 

It is worth mentioning that the values of $F_{s,u}$, $P_{s,u}$, and $Q_{s,u,l}$ must be computed prior to updating $\tilde{\mathbf{W}}_s$. These intermediate parameters are determined as 
\begin{subequations}\label{inter_var_per_sat}
\begin{align}
F_{s,u} &= F_u - \bar{\alpha}_{s,u} \mathbf{g}_{s,u}^{\mathsf{T}} \tilde{\mathbf{w}}_{s,u}, \\
P_{s,u} &= P_u - \beta_{s,u} \left| \mathbf{g}_{s,u}^{\mathsf{T}} \tilde{\mathbf{w}}_{s,u} \right|^2,\\
Q_{s,u,l} &= Q_{u,l} - \gamma_{s,u} \left| \mathbf{g}_{s,u}^{\mathsf{T}} \tilde{\mathbf{w}}_{s,l} \right|^2.
\end{align}
\end{subequations}

Given the updated $\mu_u$ and $\nu_u$, the local subproblem for optimizing $\tilde{\mathbf{W}}_s$ can be formulated as
\begin{subequations}\label{dist_bf_opt}
\begin{align}  
\mathop {\min }\limits_{\tilde{\mathbf{W}}_s} \;\;\; &-2\Omega_s + \sum_{u=1}^{U} \nu_u \left|\mu_u\right|^2 Z_u \notag \\ 
\text{{\rm s.t.}} \;\;\;  
& \left(1- \delta_{s,u}\right)\tilde{\mathbf{w}}_{s,u} = \mathbf{0}_{ N_{\text{RF}}},\;\forall u,\label{new_prob_schedule}\\
&\left\| \mathbf{F}_s \tilde{\mathbf{W}}_s\right\|^2_{\text{F}}  \le \frac{P_s}{K}.\label{new_prob_pow_bud}
\end{align}    
\end{subequations}
It is straightforward to see that \eqref{dist_bf_opt} is a convex \ac{qcqp}, which can be efficiently solved using CVX.

\renewcommand{\algorithmicrequire}{\textbf{Input:}}
\renewcommand{\algorithmicensure}{\textbf{Output:}}
\begin{algorithm}[t]
\caption{\ac{wmmse}-Based Networked \ac{leo} Satellite Distributed Beamforming over a Ring Topology-Based \acp{isl}}
\label{wmmse_ring}
\begin{algorithmic}[1]
\State \textbf{Initialize}: {$\tilde{\mathbf{W}}_s[k]$, $F_u[k]$, $P_u[k]$, $Q_{u,l}[k]$, $\forall s,u,l,k$;}
\For {$k = 1 : K$}
  \Repeat
    \For {$s = 1 : S$}
      \State Receive $F_u[k]$, $P_u[k]$, and $Q_{u,l}[k]$ from the previous satellite;
      \State Update local variables $F_{s,u}[k]$, $P_{s,u}[k]$, and $Q_{s,u,l}[k]$ via \eqref{inter_var_per_sat};
      \State Update $\mu_{u}[k]$ and $\nu_{u}[k]$ using \eqref{update_mu_and_nu_new};
      \State Update $\tilde{\mathbf{W}}_s[k]$ by solving \eqref{dist_bf_opt} via CVX;
      \State Update $F_u[k]$, $P_u[k]$, and $Q_{u,l}[k]$ via \eqref{inter_var} and relay them to the next satellite;
    \EndFor
  \Until {the relative reduction in the objective value falls below a predefined threshold or a maximum number of iterations is reached;}
\EndFor
\State \textbf{Output}: {$\tilde{\mathbf{W}}_s[k]$, $\forall s,k$.}
\end{algorithmic}
\end{algorithm}

\begin{figure}[t] 
		\centering
		\includegraphics[width=0.75\linewidth]{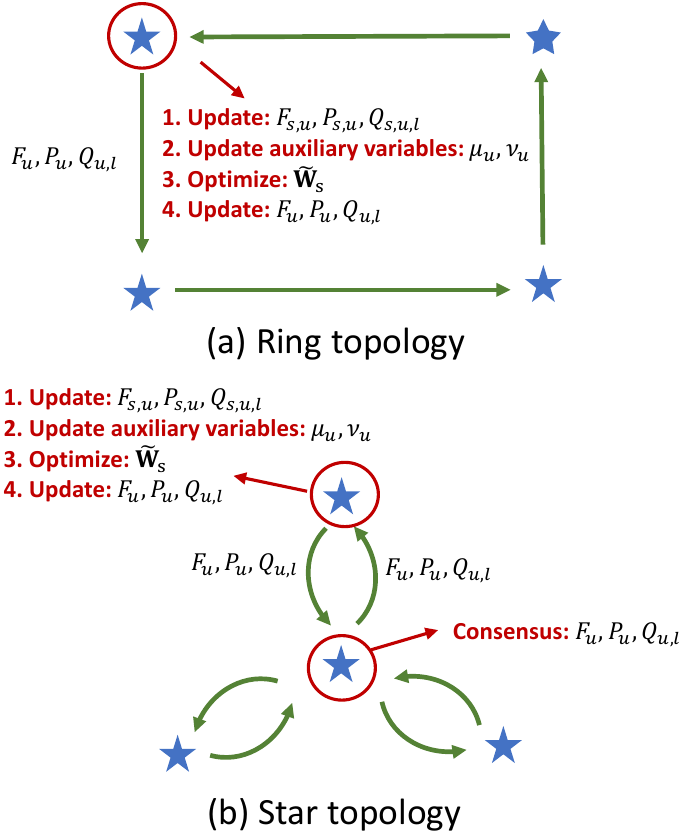}
		\caption{Workflow under different \ac{isl} topologies: (a) Ring topology and (b) Star topology.}
		\label{data_flow}
\end{figure} 

\subsection{Overall Workflow}
Based on the above discussion, the decentralized deployment of the \ac{wmmse}-based algorithm for networked \ac{leo} satellite distributed beamforming over the Ring topology-based \acp{isl} can be summarized as follows. Without loss of generality, the intermediate parameters $F_{u}$, $P_{u}$, and $Q_{u,l}$ are relayed from the $(s-1)$-th satellite to the $s$-th satellite. After reaching the $S$-th satellite, these variables are forwarded back to the first satellite, completing the ring. At each satellite, as illustrated in Fig.~\ref{data_flow}(a), once the intermediate parameters $F_{u}$, $P_{u}$, and $Q_{u,l}$ are received, the variables $F_{s,u}$, $P_{s,u}$, and $Q_{s,u,l}$ are first updated according to \eqref{inter_var_per_sat}, before the previous beamformer $\tilde{\mathbf{W}}_s$ is overwritten. Then, the auxiliary variables $\mu_u$ and $\nu_u$, as well as the beamformer $\tilde{\mathbf{W}}_s$, are sequentially updated according to \eqref{update_mu_and_nu_new} and \eqref{dist_bf_opt}. Subsequently, the updated $F_{u}$, $P_{u}$, and $Q_{u,l}$ are computed using \eqref{inter_var} and relayed to the next satellite in the ring dataflow.

The entire process, following the redemption of the subcarrier index $k$, is summarized in Algorithm \ref{wmmse_ring}. The convergence of Algorithm \ref{wmmse_ring} directly follows from the use of the \ac{wmmse} framework. As a decentralized algorithm, the per-satellite computational complexity in each iteration is dominated by solving \eqref{dist_bf_opt}, which, due to its \ac{qcqp} nature, has a complexity of $\mathcal{O}((N_{\text{RF}}T)^3)$. Note that the use of steering-vector-based analog beamforming reduces the optimization dimension from the order of the number of antennas to that of the \acp{rfc}, which are typically much fewer. This leads to a substantial complexity reduction, making the proposed design more practical and implementation-friendly for resource-constrained \ac{leo} satellites.

\begin{remark}\label{remark-ring}
We conclude that the algorithmic complexity is offloaded from the \ac{cpu} to the participating \ac{leo} satellites by the proposed distributed beamforming scheme in Algorithm \ref{wmmse_ring}, significantly enhancing the scalability of satellite cooperation. However, despite the advantage in scalability, one potential drawback of Algorithm \ref{wmmse_ring} is that the decentralized deployment incurs additional \ac{isl} backhaul overhead for relaying the intermediate parameters $F_u[k]$, $P_u[k]$, and $Q_{u,l}[k]$ ($\forall s,u,l,k$), compared to the centralized deployment in Algorithm~\ref{wmmse_cent}. The total dimensionality of this per-iteration \ac{isl} backhaul overhead is given by $D_{\text{Ring}} = K(U^2 + 2U)$. Note also that due to the iterative nature of the \ac{wmmse}-based solution, the number of iterations required for convergence may result in longer latency compared to the centralized deployment, owing to the propagation delay associated with relaying intermediate parameters via \acp{isl}.    
\end{remark}

\section{Distributed Beamforming Optimization Over Star Topology-Based ISLs}\label{sec-5}
In this section, we propose a networked \ac{leo} satellite distributed beamforming approach by adapting the \ac{wmmse} optimization framework to the Star topology-based \acp{isl}. The exchange of intermediate optimization parameters among the involved satellites is facilitated by a star-structured workflow. We begin by highlighting the unique requirement of distributed beamforming optimization via Star topology-based \acp{isl}, namely, achieving \emph{consensus}, followed by a recap of the overall protocol.

\subsection{Achieving Consensus}
In contrast to the Ring topology where each involved satellite plays a symmetric role and local processing proceeds unidirectionally through the network, the Star topology features a central satellite that exchanges information bidirectionally with the edge satellites (see Fig. \ref{data_flow}). Thus, in the distributed beamforming framework based on the Star topology, the edge satellites, upon receiving the initial intermediate parameters $F_u$, $P_u$, and $Q_{u,l}$ from the central satellite, simultaneously carry out local updates of the variables $F_{s,u}$, $P_{s,u}$, and $Q_{s,u,l}$, the auxiliary variables $\mu_u$ and $\nu_u$, the beamformer $\tilde{\mathbf{W}}_s$, and the intermediate parameters $F_u$, $P_u$, and $Q_{u,l}$, instead of performing local update sequentially. Note that this local update process is also performed at the central satellite, concurrently with the updates at the edge satellites. Afterwards, $F_u$, $P_u$, and $Q_{u,l}$ updated at the edge satellites are fed back to the central satellite. Since $F_u$, $P_u$, and $Q_{u,l}$ are updated in parallel by different edge satellites as well as the central satellite, discrepancies may arise in their values. Therefore, consensus over these intermediate parameters must be achieved at the central satellite before proceeding to the next iteration.

Let $\mathbf{f} = [F_{1},\ldots,F_{U}]^{\mathsf{T}}$, $\mathbf{p} = [P_{1},\ldots,P_{U}]^{\mathsf{T}}$, and $\mathbf{Q} \in \mathbb{R}^{U \times U}$ be the matrix whose $(u,l)$-th element is $Q_{u,l}$. Note that, to maintain generality, we do not explicitly specify which satellite serves as the central satellite in the network. To ensure consistency, the intermediate parameters must reach consensus under a unified objective within the overall optimization framework, i.e., maximizing $\sum_{u=1}^{U} (\ln \nu_{u} - \nu_{u} \Upsilon_{u})$, which is equivalent to minimizing $\sum_{u=1}^{U} \nu_{u} \Upsilon_{u}$ for a given $\nu_{u}$. 
Recall that $\Upsilon_{u}$ is rewritten as a function of $\mathbf{f}$, $\mathbf{p}$, and $\mathbf{Q}$ in \eqref{gamma_express}. 
The corresponding optimization problem to determine $\mathbf{f}$, $\mathbf{p}$, and $\mathbf{Q}$, while achieving consensus, is formulated as
\begin{subequations}\label{achieve_consensus} 
\begin{align} 
\mathop {\min }\limits_{\mathbf{f},\mathbf{p},\mathbf{Q}} \;\;\; &\sum_{u=1}^{U}\nu_u \Upsilon_u \notag \\ 
\text{{\rm s.t.}} \;\;\;  
& \mathbf{f} = \mathbf{f}_{s},\;\; \mathbf{p} = \mathbf{p}_{s},\;\; \mathbf{Q} = \mathbf{Q}_{s},\;\; \forall s,\\
& \mathbf{p} \ge \mathbf{0}_U,\;\; \text{vec}\left(\mathbf{Q}\right) \ge \mathbf{0}_{U^2},\label{pos_inter_par}
\end{align}    
\end{subequations}
where $\mathbf{f}_s$, $\mathbf{p}_s$, and $\mathbf{Q}_s$ are parameters updated by the $s$-th satellite, and constraint \eqref{pos_inter_par} is inherently required by the definitions in \eqref{inter_var_P} and \eqref{inter_var_Q}.

\renewcommand{\algorithmicrequire}{\textbf{Input:}}
\renewcommand{\algorithmicensure}{\textbf{Output:}}
\begin{algorithm}[t]
\caption{PDD-Based Consensus Algorithm}
\label{pdd-overall}
\begin{algorithmic}[1]
\State \textbf{Initialize}: $\boldsymbol{\Theta}_s=\boldsymbol{0}_{(U+2) \times U},\forall s$, $\rho$, $\delta$, $j = 1$;
\Repeat
\State {Update $\left(\mathbf{f},\mathbf{p},\mathbf{Q}\right)$ by solving \eqref{achieve_consensus_alp};}
\If {$h\left(\boldsymbol{\Gamma}\right) \le \zeta^{\left(j\right)}$}
\State {$\boldsymbol{\Theta}_s = \boldsymbol{\Theta}_s + \rho \left(\boldsymbol{\Gamma} - \boldsymbol{\Gamma}_s \right),\forall s$;}
\Else 
\State {$\rho = \delta \rho$;}
\EndIf
\State {$j = j + 1$}
\Until {$h\left(\boldsymbol{\Gamma}\right)$ is below a specified threshold or a maximum number of iterations is reached;}
\State \textbf{Output}: {$\mathbf{f}$, $\mathbf{p}$, $\mathbf{Q}$.} 
\end{algorithmic}
\end{algorithm}

The above convex problem with equality constraints can be efficiently solved using the \ac{pdd} framework~\cite{qingjiang2020tsp,zack2025wcl}, which relaxes complex constraints by adding them as penalty terms at the objective function. It iteratively updates both primal and dual variables to achieve balance between feasibility and optimality while keeping the problem tractable. Specifically, a double-loop structure is employed: The inner loop optimizes the augmented Lagrangian problem (ALP) of the original formulation, while the outer loop updates the Lagrangian dual variables and penalty parameters\cite{qingjiang2020tsp}. 

For the inner loop of the \ac{pdd} framework, the ALP corresponding to \eqref{achieve_consensus} is formulated as
\begin{align}\label{achieve_consensus_alp}  
\mathop {\min }\limits_{\mathbf{f},\mathbf{p},\mathbf{Q}} \;\;\; &\sum_{u=1}^{U}\nu_u \Upsilon_u + \frac{\rho}{2}\sum_{s=1}^{S}\left\| \boldsymbol{\Gamma} - \boldsymbol{\Gamma}_s  +\frac{1}{\rho}\boldsymbol{\Theta}_s \right\|_{\text{F}}^2 \notag \\ 
\text{{\rm s.t.}} \;\;\;  
& \mathbf{p} \ge \mathbf{0}_U,\;\; \text{vec}\left(\mathbf{Q}\right) \ge \mathbf{0}_{U^2},
\end{align}    
where $\boldsymbol{\Gamma} = [\mathbf{f},\mathbf{p},\mathbf{Q}^{\mathsf{T}}]^{\mathsf{T}}$ and $\boldsymbol{\Gamma}_s = [\mathbf{f}_s,\mathbf{p}_s,\mathbf{Q}_s^{\mathsf{T}}]^{\mathsf{T}}$. Here, $\boldsymbol{\Theta}_s \in \mathbb{C}^{(U+2) \times U}$ and $\rho$ denote the Lagrangian dual variable and penalty parameter, respectively. Note that the inner ALP problem in \eqref{achieve_consensus_alp} is a convex \ac{qcqp}, and thus its optimal solution can be efficiently obtained using CVX directly, without the need for an inner loop by block coordinate descent \cite{qingjiang2020tsp}.

Following the principle of the \ac{pdd} framework~\cite{qingjiang2020tsp}, to ensure convergence, we define the violation function as $h(\boldsymbol{\Gamma}) = \min\;(h_1(\boldsymbol{\Gamma}),\ldots,h_S(\boldsymbol{\Gamma}))$, where $h_s(\boldsymbol{\Gamma}) = \|\boldsymbol{\Gamma} - \boldsymbol{\Gamma}_s  +\frac{1}{\rho}\boldsymbol{\Theta}_s\|_{\infty}$. The detailed steps for updating the Lagrangian dual variable $\boldsymbol{\Theta}_s$ and the penalty parameter $\rho$ in the outer loop, along with the overall procedure for solving \eqref{achieve_consensus} using the \ac{pdd} approach, are summarized in Algorithm~\ref{pdd-overall}. Here, $\delta > 1$ is a constant used to increase the penalty parameter when necessary, while $\zeta^{(j)}$ denotes an empirically defined sequence that asymptotically approaches zero. Specifically, we define $\zeta^{(j)} = q h^{(j-1)}(\boldsymbol{\Gamma})$, where $q \in (0,1)$ is an attenuation constant, and $h^{(j-1)}(\boldsymbol{\Gamma})$ is the value of $h(\boldsymbol{\Gamma})$ at the $(j-1)$-th iteration.

\renewcommand{\algorithmicrequire}{\textbf{Input:}}
\renewcommand{\algorithmicensure}{\textbf{Output:}}
\begin{algorithm}[t]
\caption{\ac{wmmse}-Based Networked \ac{leo} Satellite Distributed Beamforming over a Star Topology-Based \acp{isl}}
\label{wmmse_star}
\begin{algorithmic}[1]
\State \textbf{Initialize}: {$\mathbf{W}_s[k]$, $F_u[k]$, $P_u[k]$, $Q_{u,l}[k]$, $\forall s,u,l,k$;}
\For {$k = 1 : K$}
  \Repeat
    \For {$s = 1 : S$}
      \State Receive $F_u[k]$, $P_u[k]$, and $Q_{u,l}[k]$ from the central satellite;
      \State Update $\mu_{u}[k]$ and $\nu_{u}[k]$ using \eqref{update_mu_and_nu_new};
      \State Update local variables $F_{s,u}[k]$, $P_{s,u}[k]$, and $Q_{s,u,l}[k]$ via \eqref{inter_var_per_sat};
      \State Update $\mathbf{W}_s[k]$ by solving \eqref{dist_bf_opt} via CVX;
      \State Update $F_u[k]$, $P_u[k]$, and $Q_{u,l}[k]$ via \eqref{inter_var} and relay them to the central satellite;
    \EndFor
    \State At the central satellite, achieve consensus over $F_u[k]$, $P_u[k]$, and $Q_{u,l}[k]$ via Algorithm~\ref{pdd-overall};
    \State Send the consensused $F_u[k]$, $P_u[k]$, and $Q_{u,l}[k]$ to all edge satellites;
  \Until{the relative reduction in the objective value falls below a predefined threshold or a maximum number of iterations is reached;}
\EndFor
\State \textbf{Output}: {$\mathbf{W}_s[k]$, $\forall s,k$.}
\end{algorithmic}
\end{algorithm}

\subsection{Overall Workflow}
Built upon the above, the decentralized deployment process of the \ac{wmmse}-based algorithm for networked \ac{leo} satellite distributed beamforming over a Star topology-based \acp{isl}, following the redemption of the subcarrier index $k$, is summarized in Algorithm~\ref{wmmse_star}. Similar to Algorithm~\ref{wmmse_ring}, the per-satellite computational complexity for the local update in each iteration is dominated by solving \eqref{dist_bf_opt}, which has a complexity of $\mathcal{O}((N_{\text{RF}}T)^3)$. At the central satellite, the additional requirement of achieving consensus introduces further complexity, primarily from solving \eqref{achieve_consensus_alp} in each iteration, with a complexity of $\mathcal{O}((U(U+2))^3)$.

\begin{remark}\label{remark-star-i}
Steps 4–11 of Algorithm~\ref{wmmse_star}, corresponding to the local beamformer updates at all satellites, are executed in parallel. In contrast, the local updates in the Ring topology are carried out sequentially. As will be shown in the simulations, this parallelism reduces the number of iterations required for convergence, potentially resulting in lower latency, especially when accounting for the propagation delay associated with intermediate parameter exchanges over \acp{isl}. The total dimensionality of the per-iteration \ac{isl} communication overhead from the perspective of an edge satellite is the same as that under the Ring topology, denoted by $D_{\text{Star}}^{\text{E}} = K(U^2 + 2U)$. However, due to parallelism, the per-iteration \ac{isl} communication overhead from the perspective of the central satellite is denoted by $D_{\text{Star}}^{\text{C}} = K(S-1)(U^2 + 2U)$. 
\end{remark}

\begin{table}[t]
    \centering
    \caption{Simulation Parameters}
    \label{simu_para}
    \begin{tabular}{@{}ll@{}}
        \toprule
        \textbf{Parameter} & \textbf{Value} \\ 
        \midrule
        Carrier frequency $f_c$ & $12.7 \text{ GHz}$ (Ku band) \\
        Subcarrier spacing $\Delta f$ & $120 \text{ KHz}$ \\
        Subcarrier number $K$ & 1024 \\
        Power budget at each \ac{leo} satellite & $50 \text{ dBm}$\\
        \ac{psd} $N_0$ & $-173.855 \text{ dBm/Hz}$ \\
        Noise figure $F$ & $10 \,\text{dB}$\\
        Number of \ac{leo} satellites $S$ & 4\\
        Number of \acp{ut} $U$ & 16\\
        Number of \ac{rfc} $N_{\text{RF}}$ & 8 \\
        Antenna number $N = N_{\text{h}} \times N_{\text{v}}$ & $16 \times  16$\\
        Antenna radiation gain $G(\theta)$ & $\sqrt{\frac{3}{4\pi}}\cos(\theta)$\cite{balanis2005antenna} \\
        \bottomrule
    \end{tabular}
\end{table}

\section{Numerical Results}\label{sec-6}
\subsection{Simulation Setting}
We model the Earth as a sphere with a radius of $6400\,\text{km}$. The region of interest, where \acp{ut} are randomly distributed, lies within a circular service area on the sphere with a radius of $200\,\text{km}$\cite{moewin2023jsac}. The participating \ac{leo} satellites are distributed within a parallel circular area, whose center is located along the extension of the line from the Earth's core to the center of the service area, at an orbital height of $500\,\text{km}$. The \ac{upa} installed on each satellite is oriented tangentially to the orbit, with its local coordinate system's $z$-axis pointing toward the Earth's core and its $x$-axis perpendicular to the orbital plane\cite{zack2025pace}. The components that constitute the large-scale path loss $\beta_{s,u}$ are generated according to \eqref{large_loss_component}, as well as the elaborations and remarks that follow. The Rician factors are randomly chosen between $15$ and $20\,\text{dB}$\cite{moewin2023jsac}. The remaining simulation parameters are listed in Table \ref{simu_para}.

\subsection{Benchmark Schemes}
We evaluate the performance of three proposed \ac{wmmse}-based distributed beamforming schemes for networked \ac{leo} satellites: one centralized scheme and two decentralized schemes operating under Ring and Star topologies of \acp{isl}, referred to as Central, Ring, and Star for brevity. For comparison, we introduce two networked \ac{leo} satellite beamforming baselines (under the same user scheduling policy):
\begin{itemize}
    \item \ac{mrt}: At each \ac{leo} satellite, the beamformer is denoted by $\zeta_s \mathbf{w}_{s,u}[k]$. If $u \in \mathcal{U}_s$, $\mathbf{w}_{s,u}[k]$ is designed to align with $\mathbf{g}_{s,u}$; otherwise, $\mathbf{w}_{s,u}[k] = \mathbf{0}_T$. Here, $\zeta_s$ is a normalization factor that ensures compliance with the power constraint. This method is also used to initialize the beamformers in the proposed centralized and decentralized \ac{wmmse}-based approaches;
    \item ZF: At each \ac{leo} satellite, if $u \in \mathcal{U}_s$, the beamformer is designed to lie in the null space of $\{\mathbf{g}_{s,j}\}_{j \in \mathcal{U}_s \setminus \{u\}}$, with a normalization factor for power constraint. Otherwise, $\mathbf{w}_{s,u}[k] = \mathbf{0}_T$.
\end{itemize}

To further highlight the advantages of cooperative multi-satellite systems, we also consider three baselines based on single-satellite service ($\text{S}^3$), where each \ac{ut} is exclusively served by the \ac{leo} satellite closest to it. Under the $\text{S}^3$ setting, we apply the \ac{wmmse}, \ac{mrt}, and ZF beamforming individually, denoted as \ac{wmmse}-$\text{S}^3$, \ac{mrt}-$\text{S}^3$, and ZF-$\text{S}^3$, respectively. Specifically, for \ac{wmmse}-$\text{S}^3$, the beamformer optimization is carried out independently at each satellite without considering potential inter-satellite interference.

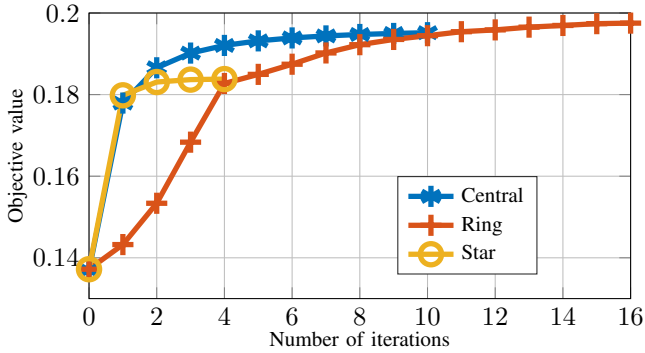
\begin{figure}[t]
\centering 
\centerline{
%
%
\definecolor{mycolor1}{rgb}{0.00000,0.44700,0.74100}%
\definecolor{mycolor2}{rgb}{0.85000,0.32500,0.09800}%
\definecolor{mycolor3}{rgb}{0.92900,0.69400,0.12500}%
\begin{tikzpicture}

\begin{axis}[%
width=72mm,
height=38mm,
at={(0mm, 0mm)},
scale only axis,
xmin=0,
xmax=16,
xlabel style={font=\color{white!15!black}, font=\footnotesize, yshift=6pt},
xlabel={Number of iterations},
ymin=0.13,
ymax=0.2,
ylabel style={font=\color{white!15!black}, font=\footnotesize, yshift=-8pt, xshift=0pt},
ylabel={Objective value},
axis background/.style={fill=white},
xmajorgrids,
ymajorgrids,
legend style={legend cell align=left, align=left, draw=white!15!black, font=\footnotesize, at={(0.828,0.429)}}
]
\addplot [color=mycolor1, line width=2.0pt, mark size=4.0pt, mark=asterisk, mark options={solid, mycolor1}]
  table[row sep=crcr]{%
0	0.136937210369209\\
1	0.177928533072948\\
2	0.186572231417575\\
3	0.190177263509174\\
4	0.192052246848345\\
5	0.193168338712563\\
6	0.193890818606967\\
7	0.194384659415282\\
8	0.194734462759118\\
9	0.194987646265142\\
10	0.195173817086098\\
};
\addlegendentry{Central}

\addplot [color=mycolor2, line width=2.0pt, mark size=4.0pt, mark=+, mark options={solid, mycolor2}]
  table[row sep=crcr]{%
0	0.137200191591212\\
1	0.143242319134745\\
2	0.153366303026887\\
3	0.168339759114153\\
4	0.182733148957787\\
5	0.184958651628485\\
6	0.1874442470218\\
7	0.190210299477945\\
8	0.19223603434018\\
9	0.193519134044665\\
10	0.194440071483043\\
11	0.195416381438894\\
12	0.195859920724456\\
13	0.196533638134165\\
14	0.196945765274482\\
15	0.197388149453255\\
16	0.197519145628714\\
};
\addlegendentry{Ring}

\addplot [color=mycolor3, line width=2.0pt, mark size=4.0pt, mark=o, mark options={solid, mycolor3}]
  table[row sep=crcr]{%
0	0.137200191591212\\
1	0.179825691923607\\
2	0.183080340137086\\
3	0.183673439364483\\
4	0.18379781732621\\
};
\addlegendentry{Star}

\end{axis}
\end{tikzpicture}%
}
\vspace{-0.2cm}
\caption{Convergent behavior of the proposed networked \ac{leo} satellite distributed beamforming schemes.}
\label{converge}
\end{figure}

\subsection{Simulation Results}
In Fig. \ref{converge}, we investigate the convergence behavior of the proposed \ac{wmmse}-based networked \ac{leo} satellite distributed beamforming approaches. We observe that, compared to Central, Ring converges to a comparable objective value, albeit at the cost of requiring more iterations. In contrast, Star converges to a lower objective value with fewer iterations, owing to its structure where edge satellites update in parallel without waiting for results from others. It is important to highlight that, unlike Central where iterations are executed internally within the \ac{cpu}, iterations of Ring and Star involve the exchange of intermediate parameters via \acp{isl}, introducing additional latency (see Remarks \ref{remark-ring} and \ref{remark-star-i}). Consequently, although Ring eventually converges to a higher value than Star, the excessive number of iterations could compromise its effectiveness due to practical coherence degradation over multiple iterations.

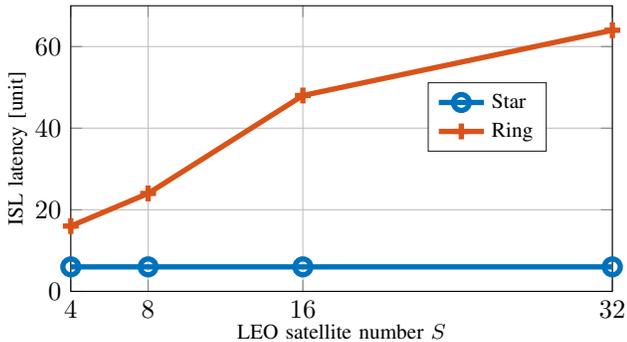
\begin{figure}[t]
\centering 
\centerline{
%
%
\definecolor{mycolor1}{rgb}{0.00000,0.44700,0.74100}%
\definecolor{mycolor2}{rgb}{0.85000,0.32500,0.09800}%
\begin{tikzpicture}

\begin{axis}[%
width=72mm,
height=38mm,
at={(0mm, 0mm)},
scale only axis,
xtick={4, 8, 16, 32}, 
xmin=4,
xmax=32,
xlabel style={font=\color{white!15!black}, font=\footnotesize, yshift=6pt},
xlabel={LEO satellite number $S$},
ymin=0,
ymax=70,
ylabel style={font=\color{white!15!black}, font=\footnotesize, yshift=-15pt, xshift=0pt},
ylabel={ISL latency [unit]},
axis background/.style={fill=white},
xmajorgrids,
ymajorgrids,
legend style={at={(0.659,0.489)}, anchor=south west, legend cell align=left, align=left, draw=white!15!black, font=\footnotesize}
]
\addplot [color=mycolor1, line width=2.0pt, mark size=3.0pt, mark=o, mark options={solid, mycolor1}]
  table[row sep=crcr]{%
4	6\\
8	6\\
16	6\\
32	6\\
};
\addlegendentry{Star}

\addplot [color=mycolor2, line width=2.0pt, mark size=3.0pt, mark=+, mark options={solid, mycolor2}]
  table[row sep=crcr]{%
4	16\\
8	24\\
16	48\\
32	64\\
};
\addlegendentry{Ring}

\end{axis}
\end{tikzpicture}%
}
\vspace{-0.2cm}
\caption{ISL latency versus \ac{leo} satellite number $S$.}
\label{delay_vs_leo}
\end{figure}

In Fig. \ref{delay_vs_leo}, we compare the incurred \ac{isl} latency as a function of the number of \ac{leo} satellites between Ring and Star. \emph{Note that, for simplicity, we ignore potential propagation latency differences across different \acp{isl}, and assume that each \ac{isl} incurs one unit of latency.} Additionally, for Ring, each algorithmic iteration incurs one unit of \ac{isl} latency (from one satellite to its immediate neighbor), whereas for Star, each iteration incurs two units of \ac{isl} latency (from the central node to an edge node and back). As observed, the total \ac{isl} latency for the Star topology remains constant as the number of participating satellites increases, while it grows linearly for the Ring topology. This behavior arises from the fundamentally different decentralization mechanisms employed by the two schemes. In the Star topology, the inherent parallelism among the edge satellites ensures that scaling the network with more satellites does not significantly increase the overall latency. In contrast, in the Ring topology, the sequential nature of local updates means that involving more satellites naturally leads to higher latency, as the total one-loop latency is proportional to the number of satellites.

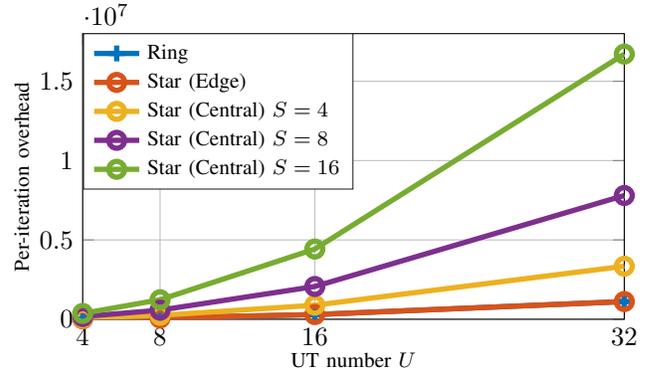
\begin{figure}[t]
\centering 
\centerline{
%
%
\definecolor{mycolor1}{rgb}{0.00000,0.44700,0.74100}%
\definecolor{mycolor2}{rgb}{0.85000,0.32500,0.09800}%
\definecolor{mycolor3}{rgb}{0.92900,0.69400,0.12500}%
\definecolor{mycolor4}{rgb}{0.49400,0.18400,0.55600}%
\definecolor{mycolor5}{rgb}{0.46600,0.67400,0.18800}%
\begin{tikzpicture}

\begin{axis}[%
width=72mm,
height=38mm,
at={(0mm, 0mm)},
scale only axis,
xtick={4, 8, 16, 32}, 
xmin=4,
xmax=32,
xlabel style={font=\color{white!15!black}, font=\footnotesize, yshift=6pt},
xlabel={UT number $U$ },
ymin=0,
ymax=18000000,
ylabel style={font=\color{white!15!black}, font=\footnotesize, yshift=-12pt, xshift=0pt},
ylabel={Per-iteration overhead},
axis background/.style={fill=white},
xmajorgrids,
ymajorgrids,
legend style={font=\footnotesize, at={(0.000609,0.454)}, anchor=south west, legend cell align=left, align=left, draw=white!15!black}
]
\addplot [color=mycolor1, line width=2.0pt, mark size=3.0pt, mark=+, mark options={solid, mycolor1}]
  table[row sep=crcr]{%
4	24576\\
8	81920\\
16	294912\\
32	1114112\\
};
\addlegendentry{Ring}

\addplot [color=mycolor2, line width=2.0pt, mark size=3.0pt, mark=o, mark options={solid, mycolor2}]
  table[row sep=crcr]{%
4	24576\\
8	81920\\
16	294912\\
32	1114112\\
};
\addlegendentry{Star (Edge)}

\addplot [color=mycolor3 , line width=2.0pt, mark size=3.0pt, mark=o, mark options={solid, mycolor3}]
  table[row sep=crcr]{%
4	73728\\
8	245760\\
16	884736\\
32	3342336\\
};
\addlegendentry{Star (Central) $S = 4$}

\addplot [color=mycolor4 , line width=2.0pt, mark size=3.0pt, mark=o, mark options={solid, mycolor4}]
  table[row sep=crcr]{%
4	172032\\
8	573440\\
16	2064384\\
32	7798784\\
};
\addlegendentry{Star (Central) $S = 8$}

\addplot [color=mycolor5 , line width=2.0pt, mark size=3.0pt, mark=o, mark options={solid, mycolor5}]
  table[row sep=crcr]{%
4	368640\\
8	1228800\\
16	4423680\\
32	16711680\\
};
\addlegendentry{Star (Central) $S = 16$}

\end{axis}
\end{tikzpicture}%
}
\vspace{-0.2cm}
\caption{Per-iteration communication overhead versus \ac{ut} number $U$.}
\label{overhead}
\end{figure}

In Fig. \ref{overhead}, we compare the per-iteration communication overhead in terms of the overall dimension of the conveyed intermediate parameters versus the number of \acp{ut} for the Ring and Star topologies. As discussed in Remark \ref{remark-star-i}, unlike the symmetric unidirectional information exchange in Ring, Star requires asymmetric bidirectional communication. From the perspective of edge satellites, the communication overhead in Star is comparable to that in Ring. However, for the central satellite in Star, the overhead scales by a factor of $S-1$. As shown, all overheads increase with the number of participating \acp{ut}. Notably, while the overhead for Ring and for the edge satellites in Star remains independent of the number of satellites, the overhead for the central satellite in Star grows as more satellites participate. 

The key insight from the above analysis is that, although the Star topology attains a substantial fraction of the centralized gain within a modest number of iterations, its signaling overhead at the central satellite scales unfavorably with the numbers of satellites and \acp{ut}. In contrast, the Ring topology eliminates the central bottleneck and can achieve a higher rate upon convergence, but it typically requires many more iterations, resulting in larger latency. These observations suggest moving beyond pure Ring or Star toward more flexible architectures. Promising directions include (i) hybrid topologies (e.g., Ring–Star) that trade central overhead for latency, (ii) fully decentralized consensus that does not rely on a central node, and (iii) exchange of dimension-compressed intermediate parameters to reduce signaling. A full exploration is beyond the scope of this paper and is left to future work.

\begin{figure}[t]
\centering 
\centerline{
%
%
\definecolor{mycolor1}{rgb}{0.00000,0.44700,0.74100}%
\definecolor{mycolor2}{rgb}{0.85000,0.32500,0.09800}%
\definecolor{mycolor3}{rgb}{0.92900,0.69400,0.12500}%
\definecolor{mycolor4}{rgb}{0.49400,0.18400,0.55600}%
\definecolor{mycolor5}{rgb}{0.46600,0.67400,0.18800}%
\definecolor{mycolor6}{rgb}{0.30100,0.74500,0.93300}%
\begin{tikzpicture}

\begin{axis}[%
name=main,
width=72mm,
height=38mm,
at={(0mm, 0mm)},
scale only axis,
xtick={16,64,256,1024},
xticklabels={16,64,256,1024},
xmin=16,
xmax=1024,
xmode = log,
log ticks with fixed point, 
xlabel style={font=\color{white!15!black}, font=\footnotesize, yshift=6pt},
xlabel={Satellite antenna number $N$},
ymin=0,
ymax=100000000,
ylabel style={font=\color{white!15!black}, font=\footnotesize, yshift=-13pt, xshift=0pt},
ylabel={Sum rate [bps]},
axis background/.style={fill=white},
xmajorgrids,
ymajorgrids,
 legend style={at={(0.0929,1.02)}, anchor=south west, legend cell align=left, align=left,
                draw=white!15!black, font=\footnotesize, legend columns=2}
]
\addplot [color=mycolor1, line width=2.0pt, mark size=4.0pt, mark=o, mark options={solid, mycolor1}]
  table[row sep=crcr]{%
16	1415193.07403703\\
64	5717589.53906151\\
256	22397706.7917162\\
1024	84939408.675031\\
4096	287613765.435057\\
};
\addlegendentry{Star (Actual)}

\addplot [color=mycolor2, dashed, line width=2.0pt, mark size=4.0pt, mark=o, mark options={solid, mycolor2}]
  table[row sep=crcr]{%
16	1418916.05452804\\
64	5749530.77178384\\
256	22598433.3592818\\
1024	83991808.2449964\\
4096	287234575.726568\\
};
\addlegendentry{Star (Nominal)}

\addplot [color=mycolor3, line width=2.0pt, mark size=4.0pt, mark=+, mark options={solid, mycolor3}]
  table[row sep=crcr]{%
16	1586233.11708108\\
64	6237359.34668845\\
256	23868998.652208\\
1024	89099625.7698072\\
4096	297136488.785923\\
};
\addlegendentry{Ring (Actual)}

\addplot [color=mycolor4, dashed, line width=2.0pt, mark size=4.0pt, mark=+, mark options={solid, mycolor4}]
  table[row sep=crcr]{%
16	1591400.42548879\\
64	6301766.9611755\\
256	24294915.3365024\\
1024	87675568.0939745\\
4096	296648721.516173\\
};
\addlegendentry{Ring (Nominal)}

\addplot [color=mycolor5, line width=2.0pt, mark size=4.0pt, mark=asterisk, mark options={solid, mycolor5}]
  table[row sep=crcr]{%
16	1601087.51398902\\
64	6288316.53307415\\
256	24008526.1002272\\
1024	89671189.0132545\\
4096	297575848.664375\\
};
\addlegendentry{Central}

\addplot [color=mycolor6, dashed, line width=2.0pt]
  table[row sep=crcr]{%
16	1744332.82661255\\
64	6704516.57471095\\
256	25315075.3543178\\
1024	91870759.4216589\\
4096	312696719.450584\\
};
\addlegendentry{Central (Instantaneous)}

\end{axis}

\begin{axis}[
  width=34mm, height=22mm,
  at={(main.north west)}, anchor=north west, xshift=7mm, yshift=-6mm,
  scale only axis,
  xmin=60, xmax=70, xmode=log, log ticks with fixed point,
  xtick={64}, xticklabels={{64}},
  ymin=5.5e6, ymax=7e6,
  title style={yshift=-2pt}, title={\scriptsize Zoom @ $N=64$},
  tick style={line width=0.3pt}, ticklabel style={/pgf/number format/fixed},
  grid=both, major grid style={opacity=0.15},
  legend style={draw=none}, 
]

\addplot [color=mycolor1, line width=2.0pt, mark size=4.0pt, mark=o, mark options={solid, mycolor1}, forget plot]
  table[row sep=crcr]{%
16	1415193.07403703\\
64	5717589.53906151\\
256	22397706.7917162\\
1024	84939408.675031\\
4096	287613765.435057\\
};

\addplot [color=mycolor2, dashed, line width=2.0pt, mark size=4.0pt, mark=o, mark options={solid, mycolor2}, forget plot]
  table[row sep=crcr]{%
16	1418916.05452804\\
64	5749530.77178384\\
256	22598433.3592818\\
1024	83991808.2449964\\
4096	287234575.726568\\
};

\addplot [color=mycolor3, line width=2.0pt, mark size=4.0pt, mark=+, mark options={solid, mycolor3}, forget plot]
  table[row sep=crcr]{%
16	1586233.11708108\\
64	6237359.34668845\\
256	23868998.652208\\
1024	89099625.7698072\\
4096	297136488.785923\\
};

\addplot [color=mycolor4, dashed, line width=2.0pt, mark size=4.0pt, mark=+, mark options={solid, mycolor4}, forget plot]
  table[row sep=crcr]{%
16	1591400.42548879\\
64	6301766.9611755\\
256	24294915.3365024\\
1024	87675568.0939745\\
4096	296648721.516173\\
};

\addplot [color=mycolor5, line width=2.0pt, mark size=4.0pt, mark=asterisk, mark options={solid, mycolor5}, forget plot]
  table[row sep=crcr]{%
16	1601087.51398902\\
64	6288316.53307415\\
256	24008526.1002272\\
1024	89671189.0132545\\
4096	297575848.664375\\
};

\addplot [color=mycolor6, dashed, line width=2.0pt, forget plot]
  table[row sep=crcr]{%
16	1744332.82661255\\
64	6704516.57471095\\
256	25315075.3543178\\
1024	91870759.4216589\\
4096	312696719.450584\\
};


\draw[red!60, dashed, line width=0.3pt]
  (main.south east)++(-2mm,2mm) -- ++(-10mm,8mm);

\end{axis}

\end{tikzpicture}%
}
\vspace{-0.2cm}
\caption{Sum rate performance versus antenna number $N$, comparing the proposed statistical \ac{csi}-based beamforming optimization (both nominal and actual) with the average sum rate achieved under instantaneous \ac{csi}-based beamforming optimization.}
\label{actual_vs_nominal_vs_inst}
\end{figure}

Fig. \ref{actual_vs_nominal_vs_inst} illustrates the sum rate performance as a function of the satellite antenna number $N$, comparing the actual and nominal rates under the proposed Ring- and Star-based decentralized optimization algorithms, thereby quantitatively validating the approximation in \eqref{rate_lb_new_b}. The nominal rate is obtained from the approximated expression in \eqref{rate_lb_new_b}, where the $\eta_{s,l}$ term representing part of the cross-satellite interference is omitted, while the actual rate is calculated from the complete expression in \eqref{rate_lb_new_a}. For reference, the centralized optimization result is also included as an upper bound. The results show that the difference between the actual and nominal rates remains marginal across the entire range of antenna numbers, confirming the validity of the approximation. This outcome is consistent with the discussion in Section~\ref{sec-4-a}: satellite communication generally operates in a noise-dominant regime due to long propagation distances, while the analog beamformer in \eqref{analog_bf} inherently suppresses interference. As a result, omitting the $\eta_{s,l}$ term, which represents part of the interference component, leads to negligible performance variation.

In addition, Fig.~\ref{actual_vs_nominal_vs_inst} also evaluates the performance gap between the proposed statistical \ac{csi}-based framework and schemes based on instantaneous \ac{csi}. Specifically, the figure presents the sum rate performance of instantaneous \ac{csi}-based distributed beamforming optimized using the WMMSE framework, with results averaged over 100 Monte Carlo trials to ensure statistical reliability. As observed, the performance loss of the proposed statistical \ac{csi}-based optimization is marginal compared to the instantaneous \ac{csi}-based benchmark. This robustness is attributed to the \ac{los}-dominant nature of \ac{leo} satellite channels, where the steering-vector-based analog beamforming stage effectively suppresses interference prior to digital optimization, thereby reducing the relative advantage of instantaneous \ac{csi} for interference management.

\begin{figure}[t]
\centering 
\centerline{
%
%
\definecolor{mycolor1}{rgb}{0.00000,0.44700,0.74100}%
\definecolor{mycolor2}{rgb}{0.85000,0.32500,0.09800}%
\definecolor{mycolor3}{rgb}{0.92900,0.69400,0.12500}%
\definecolor{mycolor4}{rgb}{0.49400,0.18400,0.55600}%
\definecolor{mycolor5}{rgb}{0.46600,0.67400,0.18800}%
\definecolor{mycolor6}{rgb}{0.30100,0.74500,0.93300}%
\definecolor{mycolor7}{rgb}{0.63500,0.07800,0.18400}%
\begin{tikzpicture}

\begin{axis}[%
width=72mm,
height=38mm,
at={(0mm, 0mm)},
scale only axis,
xmode=log,
xmin=1,
xmax=100000,
xminorticks=true,
xlabel style={font=\color{white!15!black}, font=\footnotesize, yshift=6pt},
xlabel={UT position error [m]},
ymin=0,
ymax=25000000,
ylabel style={font=\color{white!15!black}, font=\footnotesize, yshift=-13pt, xshift=0pt},
ylabel={Sum rate [bps]},
axis background/.style={fill=white},
xmajorgrids,
ymajorgrids,
yminorgrids,
legend style={at={(0.0929,1.04)}, anchor=south west, legend cell align=left, align=left, draw=white!15!black, font=\footnotesize, legend columns = 4}
]
\addplot [color=mycolor1, line width=2.0pt, mark size=3.0pt, mark=o, mark options={solid, mycolor1}]
  table[row sep=crcr]{%
1	22397687.2650247\\
10	22397511.4507068\\
100	22395744.8354825\\
1000	22377237.1243228\\
10000	22100050.2984247\\
100000	12892226.014274\\
};
\addlegendentry{Star}

\addplot [color=mycolor2, line width=2.0pt, mark size=3.0pt, mark=+, mark options={solid, mycolor2}]
  table[row sep=crcr]{%
1	23868994.753159\\
10	23868959.5892362\\
100	23868600.6312181\\
1000	23864280.9878173\\
10000	23742799.2593938\\
100000	13762249.5121818\\
};
\addlegendentry{Ring}

\addplot [color=mycolor3, line width=2.0pt, mark size=3.0pt, mark=asterisk, mark options={solid, mycolor3}]
  table[row sep=crcr]{%
1	24008527.4782753\\
10	24008514.4690547\\
100	24008409.935297\\
1000	24006606.8299553\\
10000	23944045.7245671\\
100000	13840223.0477853\\
};
\addlegendentry{Central}

\addplot [color=mycolor4, line width=2.0pt, mark size=3.0pt, mark=triangle, mark options={solid, mycolor4}]
  table[row sep=crcr]{%
1	16831040.7305795\\
10	16830659.6476785\\
100	16826823.0791557\\
1000	16785909.8152202\\
10000	16156976.0107239\\
100000	6316181.17668172\\
};
\addlegendentry{MRT}

\addplot [color=mycolor5, line width=2.0pt, mark size=3.0pt, mark=diamond, mark options={solid, mycolor5}]
  table[row sep=crcr]{%
1	8907993.50290769\\
10	8908047.46522872\\
100	8908552.30182711\\
1000	8910120.90969925\\
10000	8580694.25485056\\
100000	65933.8346347123\\
};
\addlegendentry{ZF}

\addplot [color=mycolor6, line width=2.0pt, mark size=3.0pt, mark=asterisk, mark options={solid, mycolor6}]
  table[row sep=crcr]{%
1	7559765.39321794\\
10	7559776.39082169\\
100	7559866.25913176\\
1000	7559208.55193632\\
10000	7441941.89666153\\
100000	3454696.32788849\\
};
\addlegendentry{WMMSE-$\text{S}^3$}

\addplot [color=mycolor7, line width=2.0pt, mark size=3.0pt, mark=triangle, mark options={solid, mycolor7}]
  table[row sep=crcr]{%
1	6721353.91172979\\
10	6721266.42696782\\
100	6720374.19235921\\
1000	6709732.10779112\\
10000	6455893.54361223\\
100000	2572785.78117106\\
};
\addlegendentry{MRT-$\text{S}^3$}

\addplot [color=mycolor1, line width=2.0pt, mark size=3.0pt, mark=diamond, mark options={solid, mycolor1}]
  table[row sep=crcr]{%
1	5374927.42876484\\
10	5374893.38342829\\
100	5374531.12637419\\
1000	5368725.45734794\\
10000	5094466.89139522\\
100000	1451501.96550284\\
};
\addlegendentry{ZF-$\text{S}^3$}

\end{axis}
\end{tikzpicture}%
}
\vspace{-0.2cm}
\caption{Sum rate under various schemes versus  \ac{ut} position error.}
\label{rate_vs_pos_error}
\end{figure}

Fig. \ref{rate_vs_pos_error} shows the sum rate performance as a function of \ac{ut} position error, evaluating the robustness of the proposed analog beamforming design under imperfect \ac{aod} estimates. The results compare the proposed schemes with benchmark designs and indicate that noticeable degradation occurs only when the position error reaches the order of 10 km. Since modern \ac{gnss} systems typically provide meter-level accuracy (except in rare cases of severe outage), such large errors are highly unlikely in practice. This robustness stems from the fact that \ac{ut}–satellite distances are on the order of 500~km, so moderate position errors translate into negligible angular mismatches. Overall, the figure confirms that the proposed analog beamforming approach remains resilient under realistic conditions with imperfect position information.

\begin{figure}[t]
\centering 
\centerline{
%
%
\definecolor{mycolor1}{rgb}{0.00000,0.44700,0.74100}%
\definecolor{mycolor2}{rgb}{0.85000,0.32500,0.09800}%
\definecolor{mycolor3}{rgb}{0.92900,0.69400,0.12500}%
\definecolor{mycolor4}{rgb}{0.49400,0.18400,0.55600}%
\definecolor{mycolor5}{rgb}{0.46600,0.67400,0.18800}%
\definecolor{mycolor6}{rgb}{0.30100,0.74500,0.93300}%
\definecolor{mycolor7}{rgb}{0.63500,0.07800,0.18400}%
\begin{tikzpicture}

\begin{axis}[%
width=72mm,
height=38mm,
at={(0mm, 0mm)},
scale only axis,
xmin=40,
xmax=50,
xlabel style={font=\color{white!15!black}, font=\footnotesize, yshift=6pt},
xlabel={Power budget $P_s$ [dBm]},
ymin=0,
ymax=25000000,
ylabel style={font=\color{white!15!black}, font=\footnotesize, yshift=-20pt, xshift=0pt},
ylabel={Sum rate [bps]},
axis background/.style={fill=white},
xmajorgrids,
ymajorgrids,
legend style={at={(0.000609,0.504)}, anchor=south west, legend cell align=left, align=left, draw=white!15!black, font=\footnotesize, legend columns = 2}
]
\addplot [color=mycolor1, line width=2.0pt, mark size=3.0pt, mark=o, mark options={solid, mycolor1}]
  table[row sep=crcr]{%
40	2332590.79657036\\
42	3686476.88955256\\
44	5816825.70734101\\
46	9155493.38256631\\
48	14356082.6328273\\
50	22384671.7658667\\
};
\addlegendentry{Star}

\addplot [color=mycolor2, line width=2.0pt, mark size=3.0pt, mark=+, mark options={solid, mycolor2}]
  table[row sep=crcr]{%
40	2542025.01509641\\
42	4011204.56592111\\
44	6313703.52172656\\
46	9899514.01216484\\
48	15430945.1738767\\
50	23846107.2676882\\
};
\addlegendentry{Ring}

\addplot [color=mycolor3, line width=2.0pt, mark size=3.0pt, mark=asterisk, mark options={solid, mycolor3}]
  table[row sep=crcr]{%
40	2558788.81025836\\
42	4037422.59894417\\
44	6358467.86186323\\
46	9968380.58257359\\
48	15532972.7098299\\
50	23982958.6435398\\
};
\addlegendentry{Central}

\addplot [color=mycolor4, line width=2.0pt, mark size=3.0pt, mark=triangle, mark options={solid, mycolor4}]
  table[row sep=crcr]{%
40	1711177.32489218\\
42	2709027.21247696\\
44	4286000.83796989\\
46	6774099.34919912\\
48	10689625.259291\\
50	16826844.4101684\\
};
\addlegendentry{MRT}

\addplot [color=mycolor5, line width=2.0pt, mark size=3.0pt, mark=diamond, mark options={solid, mycolor5}]
  table[row sep=crcr]{%
40	894192.528852917\\
42	1416841.81302454\\
44	2244645.01091319\\
46	3555270.73191358\\
48	5629082.39091216\\
50	8907379.46736664\\
};
\addlegendentry{ZF}

\addplot [color=mycolor6, dashed, line width=2.0pt, mark size=3.0pt, mark=asterisk, mark options={solid, mycolor6}]
  table[row sep=crcr]{%
40	759862.741671874\\
42	1203931.54848262\\
44	1907173.04776964\\
46	3020337.65043078\\
48	4778930.61508481\\
50	7559763.91715191\\
};
\addlegendentry{WMMSE-$\text{S}^3$}

\addplot [color=mycolor7, dashed, line width=2.0pt, mark size=3.0pt, mark=triangle, mark options={solid, mycolor7}]
  table[row sep=crcr]{%
40	675435.615720783\\
42	1070151.69649217\\
44	1695218.85885841\\
46	2684590.66890669\\
48	4249401.33331365\\
50	6721363.61272432\\
};
\addlegendentry{MRT-$\text{S}^3$}

\addplot [color=mycolor1, dashed, line width=2.0pt, mark size=3.0pt, mark=diamond, mark options={solid, mycolor1}]
  table[row sep=crcr]{%
40	538568.568215976\\
42	853462.550328867\\
44	1352368.05002634\\
46	2142658.86662761\\
48	3394129.41391723\\
50	5374931.1871126\\
};
\addlegendentry{ZF-$\text{S}^3$}

\end{axis}
\end{tikzpicture}%
}
\vspace{-0.2cm}
\caption{Sum rate under various schemes versus power budget $P_s$.}
\label{rate_vs_pow}
\end{figure}

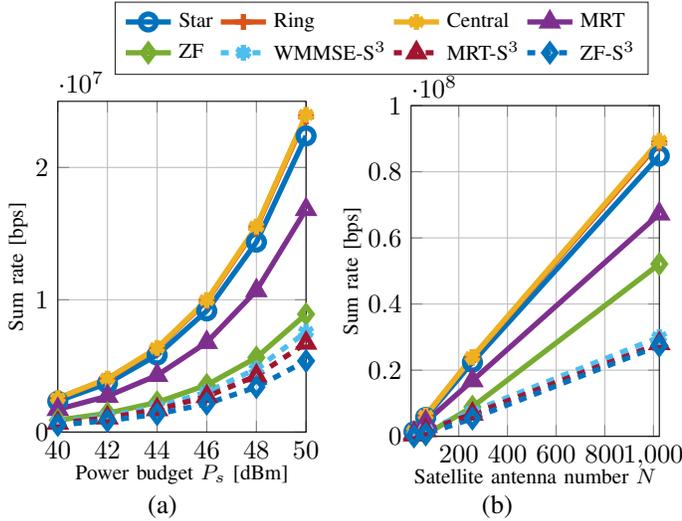
\begin{figure}[t]
\centering 
\centerline{
%
%
\definecolor{mycolor1}{rgb}{0.00000,0.44700,0.74100}%
\definecolor{mycolor2}{rgb}{0.85000,0.32500,0.09800}%
\definecolor{mycolor3}{rgb}{0.92900,0.69400,0.12500}%
\definecolor{mycolor4}{rgb}{0.49400,0.18400,0.55600}%
\definecolor{mycolor5}{rgb}{0.46600,0.67400,0.18800}%
\definecolor{mycolor6}{rgb}{0.30100,0.74500,0.93300}%
\definecolor{mycolor7}{rgb}{0.63500,0.07800,0.18400}%
%

\begin{tikzpicture}[font=\footnotesize]

\begin{axis}[
  name=main,
  width=72mm,height=38mm,
  at={(0mm,0mm)}, scale only axis,
  xtick={16,64,256,1024}, xticklabels={16,64,256,1024},
  xmin=16, xmax=1024, xmode=log, log ticks with fixed point,
  xlabel style={font=\color{white!15!black}, font=\footnotesize, yshift=6pt},
  xlabel={Satellite antenna number $N$},
  ymin=0, ymax=100000000,
  ylabel style={font=\color{white!15!black}, font=\footnotesize, yshift=-13pt, xshift=0pt},
  ylabel={Sum rate [bps]},
  axis background/.style={fill=white},
  xmajorgrids, ymajorgrids,
  legend style={at={(0.0929,1.02)}, anchor=south west, legend cell align=left, align=left,
                draw=white!15!black, font=\footnotesize, legend columns=4}
]

\addplot [color=mycolor1, line width=2.0pt, mark size=3.0pt, mark=o, mark options={solid, mycolor1}]
table[row sep=crcr]{%
16 1415147.02192808\\ 64 5716784.58149093\\ 256 22384671.7658667\\ 1024 84767415.3435186\\ 4096 286346637.656655\\};
\addlegendentry{Star}

\addplot [color=mycolor2, line width=2.0pt, mark size=3.0pt, mark=+, mark options={solid, mycolor2}]
table[row sep=crcr]{%
16 1586129.11934682\\ 64 6235757.24920968\\ 256 23846107.2676882\\ 1024 88700705.6928174\\ 4096 294742829.807969\\};
\addlegendentry{Ring}

\addplot [color=mycolor3, line width=2.0pt, mark size=3.0pt, mark=asterisk, mark options={solid, mycolor3}]
table[row sep=crcr]{%
16 1600973.3053868\\ 64 6286551.29377918\\ 256 23982958.6435398\\ 1024 89369669.6763426\\ 4096 295737448.595019\\};
\addlegendentry{Central}

\addplot [color=mycolor4, line width=2.0pt, mark size=3.0pt, mark=triangle, mark options={solid, mycolor4}]
table[row sep=crcr]{%
16 1002607.98089882\\ 64 4348984.39096697\\ 256 16826844.4101684\\ 1024 67249478.4881927\\ 4096 248055405.62283\\};
\addlegendentry{MRT}

\addplot [color=mycolor5, line width=2.0pt, mark size=3.0pt, mark=diamond, mark options={solid, mycolor5}]
table[row sep=crcr]{%
16 11189.743693808\\ 64 528939.295898289\\ 256 8907379.46736664\\ 1024 52085670.9604737\\ 4096 246998944.848006\\};
\addlegendentry{ZF}

\addplot [color=mycolor6, dashed, line width=2.0pt, mark size=3.0pt, mark=asterisk, mark options={solid, mycolor6}]
table[row sep=crcr]{%
16 474845.040138616\\ 64 1898262.1127834\\ 256 7559763.91715191\\ 1024 29969296.7942141\\ 4096 116822190.506657\\};
\addlegendentry{WMMSE-$\text{S}^3$}

\addplot [color=mycolor7, dashed, line width=2.0pt, mark size=3.0pt, mark=triangle, mark options={solid, mycolor7}]
table[row sep=crcr]{%
16 432230.482805742\\ 64 1745776.18905593\\ 256 6721363.61272432\\ 1024 28040772.0675182\\ 4096 114541180.441137\\};
\addlegendentry{MRT-$\text{S}^3$}

\addplot [color=mycolor1, dashed, line width=2.0pt, mark size=3.0pt, mark=diamond, mark options={solid, mycolor1}]
table[row sep=crcr]{%
16 35614.2373671116\\ 64 715569.46814114\\ 256 5374931.1871126\\ 1024 27430381.8934673\\ 4096 114536058.355929\\};
\addlegendentry{ZF-$\text{S}^3$}


\end{axis}

\begin{axis}[
  width=34mm, height=22mm,
  at={(main.north west)}, anchor=north west, xshift=4mm, yshift=-6mm,
  scale only axis,
  xmin=50, xmax=85, xmode=log, log ticks with fixed point,
  xtick={64}, xticklabels={{64}},
  ymin=0e6, ymax=7e6,
  title style={yshift=-2pt}, title={\scriptsize Zoom @ $N=64$},
  tick style={line width=0.3pt}, ticklabel style={/pgf/number format/fixed},
  grid=both, major grid style={opacity=0.15},
  legend style={draw=none}, 
]

\addplot [color=mycolor1, line width=2.0pt, mark size=3.0pt, mark=o, mark options={solid, mycolor1}, forget plot]
table[row sep=crcr]{%
16 1415147.02192808\\ 64 5716784.58149093\\ 256 22384671.7658667\\ 1024 84767415.3435186\\ 4096 286346637.656655\\};

\addplot [color=mycolor2, line width=2.0pt, mark size=3.0pt, mark=+, mark options={solid, mycolor2}, forget plot]
table[row sep=crcr]{%
16 1586129.11934682\\ 64 6235757.24920968\\ 256 23846107.2676882\\ 1024 88700705.6928174\\ 4096 294742829.807969\\};

\addplot [color=mycolor3, line width=2.0pt, mark size=3.0pt, mark=asterisk, mark options={solid, mycolor3}, forget plot]
table[row sep=crcr]{%
16 1600973.3053868\\ 64 6286551.29377918\\ 256 23982958.6435398\\ 1024 89369669.6763426\\ 4096 295737448.595019\\};

\addplot [color=mycolor4, line width=2.0pt, mark size=3.0pt, mark=triangle, mark options={solid, mycolor4}, forget plot]
table[row sep=crcr]{%
16 1002607.98089882\\ 64 4348984.39096697\\ 256 16826844.4101684\\ 1024 67249478.4881927\\ 4096 248055405.62283\\};

\addplot [color=mycolor5, line width=2.0pt, mark size=3.0pt, mark=diamond, mark options={solid, mycolor5}, forget plot]
table[row sep=crcr]{%
16 11189.743693808\\ 64 528939.295898289\\ 256 8907379.46736664\\ 1024 52085670.9604737\\ 4096 246998944.848006\\};

\addplot [color=mycolor6, dashed, line width=2.0pt, mark size=3.0pt, mark=asterisk, mark options={solid, mycolor6}, forget plot]
table[row sep=crcr]{%
16 474845.040138616\\ 64 1898262.1127834\\ 256 7559763.91715191\\ 1024 29969296.7942141\\ 4096 116822190.506657\\};

\addplot [color=mycolor7, dashed, line width=2.0pt, mark size=3.0pt, mark=triangle, mark options={solid, mycolor7}, forget plot]
table[row sep=crcr]{%
16 432230.482805742\\ 64 1745776.18905593\\ 256 6721363.61272432\\ 1024 28040772.0675182\\ 4096 114541180.441137\\};

\addplot [color=mycolor1, dashed, line width=2.0pt, mark size=3.0pt, mark=diamond, mark options={solid, mycolor1}, forget plot]
table[row sep=crcr]{%
16 35614.2373671116\\ 64 715569.46814114\\ 256 5374931.1871126\\ 1024 27430381.8934673\\ 4096 114536058.355929\\};


\draw[red!60, dashed, line width=0.3pt]
  (main.south east)++(-2mm,2mm) -- ++(-10mm,8mm);

\end{axis}

\end{tikzpicture}}
\vspace{-0.2cm}
\caption{Sum rate under various schemes versus antenna number $N$.}
\label{rate_vs_ant}
\end{figure}

In Figs. \ref{rate_vs_pow} and \ref{rate_vs_ant}, we compare the sum rate of various schemes as functions of the power budget and the number of antennas, respectively. It is observed that among the networked \ac{leo} satellite distributed beamforming approaches, the proposed \ac{wmmse}-based methods achieve substantially higher sum rate compared to the heuristic strategies (i.e., \ac{mrt} and ZF), demonstrating the effectiveness of the proposed optimization schemes. Within the \ac{wmmse}-based designs, Ring yields similar performance compared to Central, while Star is slightly outperformed by Central. Moreover, the networked \ac{leo} distributed beamforming schemes significantly outperform the baseline approach, where each \ac{ut} is individually served by a single satellite (i.e., the schemes with the suffix $\text{S}^3$). This performance gap widens as the power budget and antenna number increase, highlighting the substantial benefits of cooperative satellite transmission in enhancing communication efficiency.

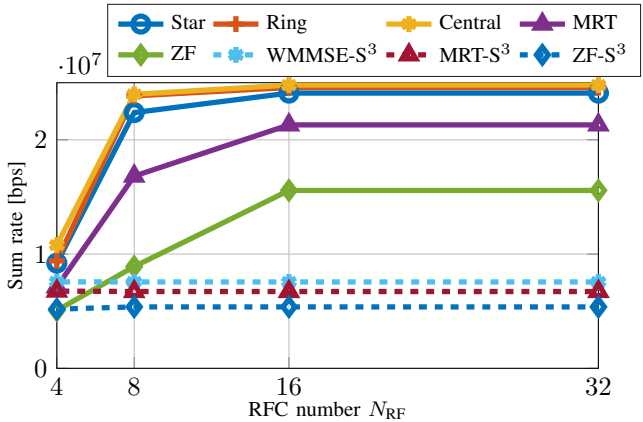
\begin{figure}[t]
\centering 
\centerline{
%
%
\definecolor{mycolor1}{rgb}{0.00000,0.44700,0.74100}%
\definecolor{mycolor2}{rgb}{0.85000,0.32500,0.09800}%
\definecolor{mycolor3}{rgb}{0.92900,0.69400,0.12500}%
\definecolor{mycolor4}{rgb}{0.49400,0.18400,0.55600}%
\definecolor{mycolor5}{rgb}{0.46600,0.67400,0.18800}%
\definecolor{mycolor6}{rgb}{0.30100,0.74500,0.93300}%
\definecolor{mycolor7}{rgb}{0.63500,0.07800,0.18400}%
\begin{tikzpicture}

\begin{axis}[%
width=72mm,
height=38mm,
at={(0mm, 0mm)},
scale only axis,
xtick={4, 8, 16, 32}, 
xmin=4,
xmax=32,
xlabel style={font=\color{white!15!black}, font=\footnotesize, yshift=6pt},
xlabel={RFC number $N_{\text{RF}}$},
ymin=0,
ymax=25000000,
ylabel style={font=\color{white!15!black}, font=\footnotesize, yshift=-20pt, xshift=0pt},
ylabel={Sum rate [bps]},
axis background/.style={fill=white},
xmajorgrids,
ymajorgrids,
legend style={at={(0.0929,1.02)}, anchor=south west, legend cell align=left, align=left, draw=white!15!black, font=\footnotesize, legend columns = 4}
]
\addplot [color=mycolor1, line width=2.0pt, mark size=3.0pt, mark=o, mark options={solid, mycolor1}]
  table[row sep=crcr]{%
1	7659059.44511991\\
2	7573555.48353462\\
4	9211934.45435594\\
8	22384671.7658667\\
16	24093650.1863332\\
32	24093650.1863255\\
64	24093650.1863255\\
};
\addlegendentry{Star}

\addplot [color=mycolor2, line width=2.0pt, mark size=3.0pt, mark=+, mark options={solid, mycolor2}]
  table[row sep=crcr]{%
1	7659059.49068419\\
2	7572506.03563589\\
4	9410229.13534039\\
8	23846107.2676882\\
16	24578063.4894918\\
32	24578063.4894929\\
64	24578063.4894929\\
};
\addlegendentry{Ring}

\addplot [color=mycolor3, line width=2.0pt, mark size=3.0pt, mark=asterisk, mark options={solid, mycolor3}]
  table[row sep=crcr]{%
1	7659059.39994664\\
2	7576518.93126808\\
4	10783858.1773791\\
8	23982958.6435398\\
16	24795190.4384973\\
32	24795190.7181379\\
64	24795190.7181379\\
};
\addlegendentry{Central}

\addplot [color=mycolor4, line width=2.0pt, mark size=3.0pt, mark=triangle, mark options={solid, mycolor4}]
  table[row sep=crcr]{%
1	7659059.50642286\\
2	6803681.79748395\\
4	7129366.22995201\\
8	16826844.4101684\\
16	21306752.111013\\
32	21306752.111013\\
64	21306752.111013\\
};
\addlegendentry{MRT}

\addplot [color=mycolor5, line width=2.0pt, mark size=3.0pt, mark=diamond, mark options={solid, mycolor5}]
  table[row sep=crcr]{%
1	7659059.50642286\\
2	3889241.86934663\\
4	5061677.00188095\\
8	8907379.46736664\\
16	15570955.164259\\
32	15570955.164259\\
64	15570955.164259\\
};
\addlegendentry{ZF}

\addplot [color=mycolor6, dashed, line width=2.0pt, mark size=3.0pt, mark=asterisk, mark options={solid, mycolor6}]
  table[row sep=crcr]{%
1	7520308.14199005\\
2	7580195.38889575\\
4	7569648.96877526\\
8	7559763.91715191\\
16	7559763.91715191\\
32	7559763.91715191\\
64	7559763.91715191\\
};
\addlegendentry{WMMSE-$\text{S}^3$}

\addplot [color=mycolor7, dashed, line width=2.0pt, mark size=3.0pt, mark=triangle, mark options={solid, mycolor7}]
  table[row sep=crcr]{%
1	7520308.20026256\\
2	7053035.50084233\\
4	6759711.03190792\\
8	6721363.61272432\\
16	6721363.61272432\\
32	6721363.61272432\\
64	6721363.61272432\\
};
\addlegendentry{MRT-$\text{S}^3$}

\addplot [color=mycolor1, dashed, line width=2.0pt, mark size=3.0pt, mark=diamond, mark options={solid, mycolor1}]
  table[row sep=crcr]{%
1	7520308.20026256\\
2	4266324.49330155\\
4	5184846.89621703\\
8	5374931.1871126\\
16	5374931.1871126\\
32	5374931.1871126\\
64	5374931.1871126\\
};
\addlegendentry{ZF-$\text{S}^3$}

\end{axis}
\end{tikzpicture}%
}
\vspace{-0.2cm}
\caption{Sum rate under various schemes versus the number \acp{rfc} $N_{\text{RF}}$.}
\label{rate_vs_rfc}
\end{figure}

In Fig. \ref{rate_vs_rfc}, we plot the sum rate under various numbers of \acp{rfc} and various schemes. The proposed \ac{wmmse}-based distributed beamforming approaches consistently outperform the baselines, with the sum rate steadily increasing as the number of \acp{rfc} grows. Nevertheless, the growth of the sum rate saturates beyond a certain \ac{rfc} threshold, reflecting the saturation of the available digital-domain \acp{dof} enabled by \ac{rfc} deployment. Across the entire comparison range, the performance gap between the decentralized Ring and Star topologies and their centralized counterpart (i.e., Central) remains mild, highlighting the robustness of the proposed decentralized networked \ac{leo} satellite distributed beamforming schemes. It is also noteworthy that increasing the number of \acp{rfc} does not substantially enhance the sum rate under the $\text{S}^3$ approaches, as fewer \acp{ut} are scheduled to each satellite (due to the non-overlapping user scheduling), causing \ac{rfc} numbers to reach abundance more quickly.

\begin{figure}[t]
\centering 
\centerline{
%
%
\definecolor{mycolor1}{rgb}{0.00000,0.44700,0.74100}%
\definecolor{mycolor2}{rgb}{0.85000,0.32500,0.09800}%
\definecolor{mycolor3}{rgb}{0.92900,0.69400,0.12500}%
\definecolor{mycolor4}{rgb}{0.49400,0.18400,0.55600}%
\definecolor{mycolor5}{rgb}{0.46600,0.67400,0.18800}%
\definecolor{mycolor6}{rgb}{0.30100,0.74500,0.93300}%
\definecolor{mycolor7}{rgb}{0.63500,0.07800,0.18400}%
\begin{tikzpicture}

\begin{axis}[%
width=72mm,
height=38mm,
at={(0mm, 0mm)},
scale only axis,
xtick={4, 8, 16, 32}, 
xmin=4,
xmax=32,
xlabel style={font=\color{white!15!black}, font=\footnotesize, yshift=6pt},
xlabel={LEO satellite number $S$},
ymin=0,
ymax=700000000,
ylabel style={font=\color{white!15!black}, font=\footnotesize, yshift=-20pt, xshift=0pt},
ylabel={Sum rate [bps]},
axis background/.style={fill=white},
xmajorgrids,
ymajorgrids,
legend style={at={(0.000609,0.504)}, anchor=south west, legend cell align=left, align=left, draw=white!15!black, font=\footnotesize, legend columns = 2}
]
\addplot [color=mycolor1, line width=2.0pt, mark size=3.0pt, mark=o, mark options={solid, mycolor1}]
  table[row sep=crcr]{%
4	26187017.9065153\\
8	80078367.4679532\\
16	226586903.137003\\
32	580799140.791709\\
};
\addlegendentry{Star}

\addplot [color=mycolor2, line width=2.0pt, mark size=3.0pt, mark=+, mark options={solid, mycolor2}]
  table[row sep=crcr]{%
4	26748224.2249015\\
8	88619865.8750947\\
16	250754915.059286\\
32	606638351.748207\\
};
\addlegendentry{Ring}

\addplot [color=mycolor3, line width=2.0pt, mark size=3.0pt, mark=asterisk, mark options={solid, mycolor3}]
  table[row sep=crcr]{%
4	27434048.9976905\\
8	91522293.0675284\\
16	253740534.233744\\
32	625909786.900351\\
};
\addlegendentry{Central}

\addplot [color=mycolor4, line width=2.0pt, mark size=3.0pt, mark=triangle, mark options={solid, mycolor4}]
  table[row sep=crcr]{%
4	22747126.2831501\\
8	64751492.3004759\\
16	171466564.310382\\
32	472698707.522535\\
};
\addlegendentry{MRT}

\addplot [color=mycolor5, line width=2.0pt, mark size=3.0pt, mark=diamond, mark options={solid, mycolor5}]
  table[row sep=crcr]{%
4	19199794.887587\\
8	66694437.3916722\\
16	193926758.101081\\
32	447775273.889856\\
};
\addlegendentry{ZF}

\addplot [color=mycolor6, dashed, line width=2.0pt, mark size=3.0pt, mark=asterisk, mark options={solid, mycolor6}]
  table[row sep=crcr]{%
4	5555930.98380753\\
8	7417060.56674254\\
16	9494929.29058925\\
32	17120813.8460533\\
};
\addlegendentry{WMMSE-$\text{S}^3$}

\addplot [color=mycolor7, dashed, line width=2.0pt, mark size=3.0pt, mark=triangle, mark options={solid, mycolor7}]
  table[row sep=crcr]{%
4	4736852.44176423\\
8	6414541.61032461\\
16	8797800.3664463\\
32	16612455.6137802\\
};
\addlegendentry{MRT-$\text{S}^3$}

\addplot [color=mycolor1, dashed, line width=2.0pt, mark size=3.0pt, mark=diamond, mark options={solid, mycolor1}]
  table[row sep=crcr]{%
4	3661321.115666\\
8	5040433.95237383\\
16	7414967.93473463\\
32	15995291.5118973\\
};
\addlegendentry{ZF-$\text{S}^3$}

\end{axis}
\end{tikzpicture}%
}
\vspace{-0.2cm}
\caption{Sum rate under various schemes versus \ac{leo} satellite number $S$.}
\label{rate_vs_leo}
\end{figure}
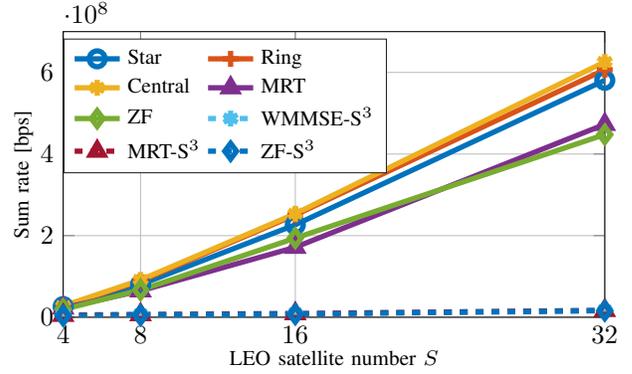

In Fig. \ref{rate_vs_leo}, we plot the sum rate as a function of the number of participating \ac{leo} satellites under various schemes. As observed, the sum rate increases significantly with the number of satellites involved in all networked \ac{leo} satellite distributed beamforming schemes, whereas it remains nearly constant under the $\text{S}^3$ approaches. This is because the networked \ac{leo} satellite schemes leverage the enlarged equivalent array aperture and enhanced cooperative array gain as more satellites participate. In contrast, under the $\text{S}^3$ approaches, each \ac{ut} is served individually by a single satellite, with satellites operating independently. This limits the ability to fully exploit the collective capabilities of the satellite constellation and prevents the improvement of spectral efficiency through cooperative transmission. These results further highlight the substantial advantages of networked \ac{leo} satellite distributed beamforming. Moreover, it is evident that the performance gap between the \ac{wmmse}-based approaches and other heuristic networked \ac{leo} satellite schemes (\ac{mrt} and ZF) widens as the number of participating satellites increases, demonstrating the effectiveness of the proposed optimization method.

\section{Conclusion}\label{sec-7}
This paper addressed a critical gap in the design of distributed beamforming schemes for networked \ac{leo} satellite systems by developing scalable solutions that operate solely on statistical \ac{csi}. We established a comprehensive system model for networked \ac{leo} satellite \ac{ofdm} downlink communications, incorporating large antenna arrays with limited \acp{rfc}, and introduced practical yet effective designs for user scheduling and analog beamforming. A closed-form lower bound on the ergodic sum rate was derived using the hardening bound, forming the basis for a sum rate maximization problem solved centrally via the \ac{wmmse} framework. Building upon this, we proposed two decentralized optimization schemes tailored to Ring and Star \acp{isl} topologies, enabling local beamformer updates and cooperative parameter exchange with distinct information flows. Extensive simulations demonstrated that the proposed distributed methods closely approached the performance of centralized method, significantly outperformed baseline schemes, and offered clear scalability advantages. Furthermore, the observed trade-offs between delay and communication overhead across topologies provided valuable insights for future research in large-scale, dynamic \ac{leo} satellite cooperation. An important direction for future work is the development of lightweight decentralized algorithms with fewer iteration requirements and adaptive mechanisms that can explicitly address satellite mobility, topology variations, and handovers, thereby enhancing the practicality of distributed optimization in dynamic \ac{leo} networks.

\section{Acknowledgment}
The authors would like to express their sincere gratitude to Dr. Seungnyun Kim of MIT and Dr. Ali Arshad Nasir of KFUPM for their insightful suggestions and fruitful discussions during the preparation of this paper.

\appendices

\bibliographystyle{IEEEtran}
\bibliography{IEEEabrv,mybib}

\begin{IEEEbiography}[{\includegraphics[width=1in,clip]{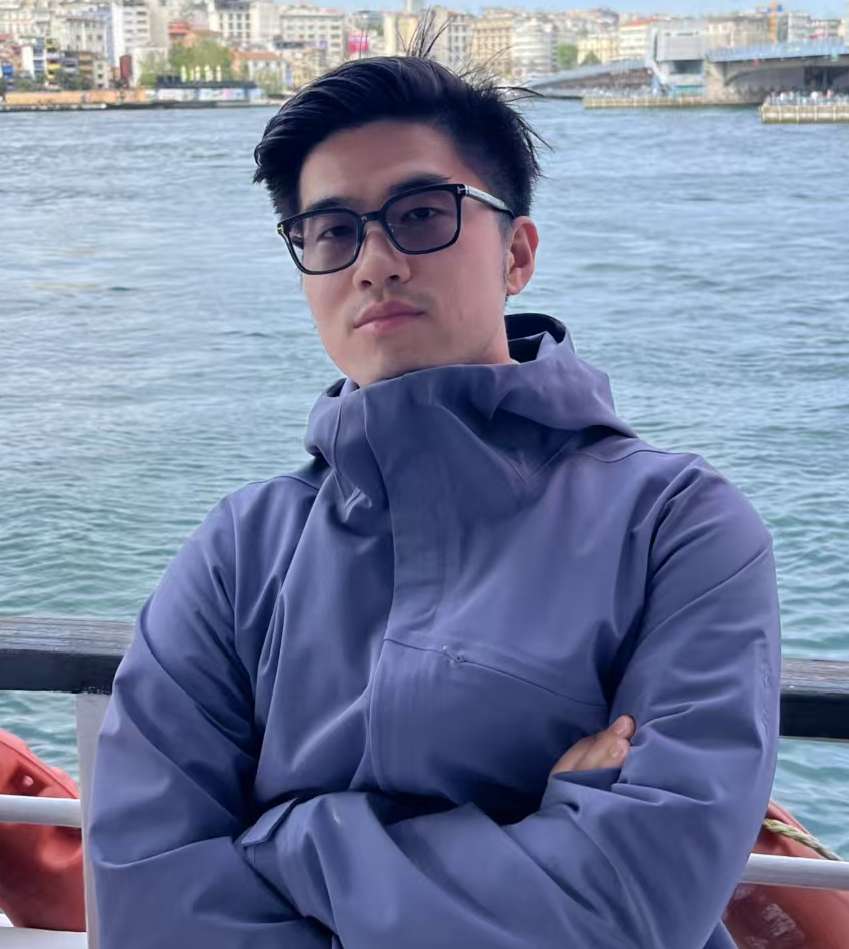}}]{Yuchen Zhang} (Member, IEEE) received the B.E. and Ph.D. degrees in communication engineering from the University of Electronic Science and Technology of China in 2018 and 2024, respectively. His Ph.D. research was supervised by Prof. Wanbin Tang, Head of the National Key Laboratory of Wireless Communications. From 2022 to 2023, he was a visiting Ph.D. student at the Weizmann Institute of Science, Israel, under the supervision of Prof. Yonina C. Eldar. He joined King Abdullah University of Science and Technology (KAUST), Saudi Arabia, in 2024 as a Postdoctoral Researcher with Prof. Tareq Y. Al-Naffouri, and was elevated to Postdoctoral Global Fellow in 2025 under the KAUST Global Fellowship program. His current research interests focus on non-terrestrial networks, particularly low-Earth-orbit satellite systems, and reconfigurable antennas for 6G-and-beyond communications, positioning, and sensing.
\end{IEEEbiography}

\begin{IEEEbiography}[{\includegraphics[width=1in,clip]{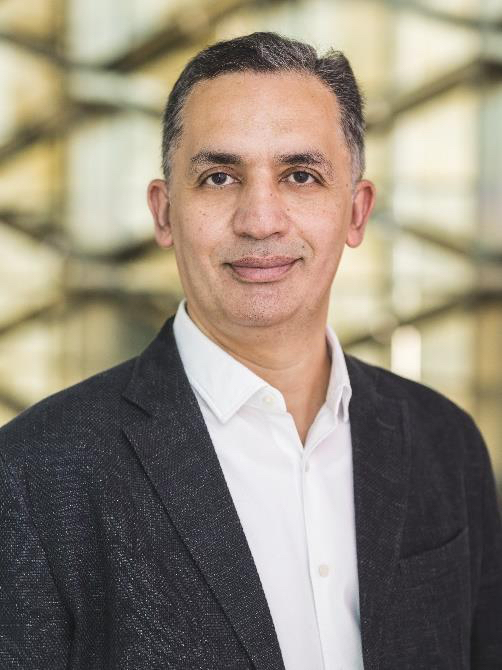}}]{Tareq Y. Al-Naffouri} (Fellow, IEEE) received the B.S. degrees in mathematics and electrical engineering (with first honors) from King Fahd University of Petroleum and Minerals, Saudi Arabia, the M.S. degree in electrical engineering from the Georgia Institute of Technology, and the Ph.D. degree in electrical engineering from Stanford University, Stanford in 2004. He was a visiting scholar at California Institute of Technology, Pasadena, CA in 2005 and summer 2006. He was a Fulbright scholar at the University of Southern California in 2008. He is currently a Professor at the Electrical Engineering Department, King Abdullah University of Science and Technology (KAUST). His research interests lie in the areas of sparse, adaptive, and statistical inference/learning and their applications to wireless communications, localization, smart cities, and smart health. He has over 370 publications in journal and conference proceedings and 24 issued/pending patents. He has won the IEEE Education Society Chapter Achievement Award (2008), Almarie Award for Innovative research in communication (2009), AbdulHameed Shoman Prize for innovative research in IoT (2022), and the Research Excellence Award, Innovation in Economies of the Future, RDIA (2025).
\end{IEEEbiography}

\end{document}